\documentclass[11pt,a4paper]{article}
\usepackage{jcappub}
\pdfoutput=1

\graphicspath{{figs/}}
\newcommand{\figh}[2]{\includegraphics[height=#1\textheight]{#2}}
\newcommand{\figw}[2]{\includegraphics[width=#1\textwidth]{#2}}

\newcommand{\uhecrs}{\mbox{UHECRs}}
\newcommand{\Cl}{\ensuremath{C_\ell}}
\newcommand{\lmax}{\ensuremath{\ell_\mathrm{max}}}
\newcommand{\Nsrc}{\ensuremath{N_\mathrm{src}}}
\newcommand{\Nside}{\ensuremath{N_\mathrm{side}}}
\newcommand{\Diso}{\ensuremath{D(\mathrm{iso})}}
\newcommand{\Dmix}{\ensuremath{D(\mathrm{mix})}}
\newcommand{\Xmax}{\ensuremath{X_\mathrm{max}}}
\newcommand{\keuso}{\mbox{K-EUSO}}

\begin{document}

\title{%
	 Identifying nearby sources of ultra-high-energy cosmic rays with deep learning
	 }

\author[a,b]{Oleg Kalashev,}
\author[a,c,d]{Maxim~Pshirkov,}
\author[e]{Mikhail~Zotov}

\affiliation[a]{Institute for Nuclear Research of the Russian Academy of
	Sciences, Moscow, 117312, Russia}
\affiliation[b]{Moscow Institute for Physics and Technology, 9 Institutskiy per.,
	Dolgoprudny, Moscow Region, 141701 Russia}
\affiliation[c]{Sternberg Astronomical Institute, Lomonosov Moscow State
	University, Moscow, 119992, Russia}
\affiliation[d]{Lebedev Physical Institute, Pushchino Radio
	Astronomy Observatory, 142290, Russia}
\affiliation[e]{Skobeltsyn Institute of Nuclear Physics,
	Lomonosov Moscow State University, Moscow, 119991, Russia}

\abstract{%
	We present a method to analyse arrival directions of
	ultra-high-energy cosmic rays (\uhecrs) using a classifier defined by
	a deep convolutional neural network trained on a HEALPix grid.  To
	illustrate a high effectiveness of the method, we employ it to
	estimate prospects of detecting a large-scale anisotropy of \uhecrs{}
	induced by a nearby source with an (orbital) detector having a
	uniform exposure of the celestial sphere and compare the results with
	our earlier calculations based on the angular power spectrum.  A
	minimal model for extragalactic cosmic rays and neutrinos by
	Kachelrie\ss, Kalashev, Ostapchenko and Semikoz (2017) is assumed for
	definiteness and nearby active galactic nuclei Centaurus~A, M82,
	NGC~253, M87 and Fornax~A are considered as possible sources of
	\uhecrs{}.  We demonstrate that the proposed method drastically
	improves sensitivity of an experiment by decreasing the minimal
	required amount of detected \uhecrs{} or the minimal detectable
	fraction of from-source events several times compared to the approach
	based on the angular power spectrum.  We also test robustness of the
	neural networks against different models of the large-scale Galactic
	magnetic fields and variations of the mass composition of \uhecrs{},
	and consider situations when there are two nearby sources or the
	dominating source is not known a~priori.  In all cases, the neural
	networks demonstrate good performance unless the test models strongly
	deviate from those used for training.  The method can be readily
	applied to the analysis of data of the Telescope Array, the Pierre
	Auger Observatory and other cosmic ray experiments.
}

\emailAdd{kalashev@inr.ac.ru}
\emailAdd{pshirkov@sai.msu.ru}
\emailAdd{zotov@eas.sinp.msu.ru}

\arxivnumber{1912.00625}
\maketitle
\flushbottom

%______________________________________________________________________
\section{Introduction}

Cosmic rays  of the highest energies ($E\gtrsim50$~EeV,
ultra-high-energy cosmic rays, \uhecrs) were first detected almost 60
years ago~\cite{1961PhRvL...6..485L} and still remain at the forefront
of the high-energy astrophysics. Due to the limited distance of CR
propagation at these energies amounting to $\sim$100~Mpc, a certain
degree of anisotropy is expected in the distribution of their arrival
directions.  The level of an anisotropy and its other properties depend
on characteristics of sources of \uhecrs, thus studying the anisotropy
is one of the key parts of this branch of astrophysics.

An analysis of an anisotropy of CR arrival directions is only feasible
when a sufficiently large amount of data is available.  Due to the
extremely low flux of \uhecrs, obtaining the required number of events
demands instruments with a very large exposure. One of possible
solutions of the problem is employing space-born detectors, which
observe UV-radiation from extensive air showers induced by \uhecrs{} in
the Earth's atmosphere.  Several missions of the kind are under
development now, such as KLYPVE-EUSO
(\keuso)~\cite{10.1093/ptep/ptx169,k-euso-ICRC2017,k-euso-ICRC2017-science,klypve-2015bullras}
and POEMMA~\cite{2019ICRC...36..378O,POEMMA}.  In our previous paper
\cite{we-jcap2019} (Paper~1 in what follows), we studied sensitivity of
these future orbital missions to a large-scale anisotropy emerging due
to the presence of a nearby source.  A particular model for cosmic rays
and neutrinos by Kachelrie\ss, Kalashev, Ostapchenko and Semikoz (KKOS
in what follows) \cite{Kachelriess:2017tvs}, which can act as a
representative of a broader class of models, was selected for the
analysis.  The model assumes that \uhecrs{} are accelerated by (a
subclass of) active galactic nuclei (AGN) with the energy spectra of
nuclei following a power-law with a rigidity-dependent cut-off after the
acceleration phase.  The model successfully reproduces the energy
spectrum of cosmic rays with energies beyond $10^{17}$~eV registered
with the Pierre Auger Observatory and the spectrum of high-energy
neutrinos registered by IceCube, as well as data on the depth of maximum
of air showers~\Xmax{} and RMS(\Xmax).  One of the consequences of the
model is the existence of a nearby (within $\sim20$~Mpc) AGN acting as a
source of UHECRs.  Presence of such an accelerator would inevitably lead
to deviations from isotropy at some level and produce detectable
imprints on the angular power spectrum (APS) of the UHECR flux providing
the fraction of nuclei arriving from the source is sufficiently high.
We considered five nearby AGN often discussed in literature as possible
sources of \uhecrs{} and demonstrated that an observation of
$\gtrsim200\text{--}300$ events with energies $\gtrsim57$~EeV will allow
detecting deviations from isotropy with a high level of statistical
significance if the fraction of events from any of these sources is
$\simeq$10--15\% of the total flux.

Use of the APS allows for a very robust approach to an analysis of
anisotropy but has certain drawbacks since some information such as the
characteristic shape and size of a region with an excessive flux of
\uhecrs{} produced by the source is only partially preserved in the APS
coefficients.  That results in a somewhat lower sensitivity to a sought
signal.  In the present paper, we searched for this signal exploiting
information about arrival directions of \uhecrs{}.
To do that, we employed a convolutional neural network (CNN) classifier.
It is known that CNNs
have demonstrated high performance in a wide range of problems related
to classification of data and pattern recognition, see,
e.g.,~\cite{2019RvMP...91d5002C}.
However, to the best of our knowledge, they have
never been employed for anisotropy studies in UHECR physics yet.

In what follows, we present a CNN trained on a HEALPix grid~\cite{healpix} and
demonstrate that it presents a real
breakthrough by reducing the number of events needed for establishing a
large-scale anisotropy produced by \uhecrs{} arriving from a nearby
source by $\sim4$ times.
Alternatively, for a fixed sample size, the CNN strongly decreases the
fraction of CRs arriving from the source necessary for a robust
detection of an anisotropy.

%______________________________________________________________________
\section{Traditional approach}
\label{sec:method}

Let us briefly remind the key points of Paper~1, which was based on
calculating the angular power spectrum of CR arrival directions mostly
following a method suggested by the IceCube and the Pierre Auger
Observatory collaborations~\cite{IceCube-aniso-2007,Auger-aniso-2017}.
We considered five nearby active galactic nuclei Centaurus~A, M82,
NGC~253, M87 and Fornax~A as possible sources of \uhecrs.  All of them
are located within a sphere with a radius of $\sim20$~Mpc with the first
three being as close as $\approx3.5\text{--}4$~Mpc from the Galaxy.  For
the aims of the analysis, we generated multiple sets of mock maps
imitating arrival directions of nuclei coming from these sources.

In a simplified approach of the KKOS model, all sources share the same
injection spectrum and composition. These properties were found from the
global fit of the CR spectrum and composition observed at Earth.
However, both composition and spectrum evolve during propagation of
nuclei in the inter-galactic media. That was taken into account with the
TransportCR code~\cite{transportcr}.  We only considered nuclei with
energies above 57~EeV, which approximately corresponds to the 100\%
efficiency threshold of the \keuso{} and POEMMA projects.  This allowed
us to assume  that extragalactic magnetic fields do not strongly deflect
the nuclei~\cite{Pshirkov:2015tua} so that \uhecrs{} arrive to the Milky
Way within $\pm1^{\circ}$ from the original direction.  Next, we
employed the CRPropa~3 code~\cite{CRPropa} to simulate propagation of
nuclei in the Galactic magnetic field (GMF), for which we assumed the
Jansson--Farrar model~\cite{JF12,JF12b}, JF12 in what follows.  All
three components of the magnetic field present in the model (the
regular, striated and turbulent ones) were utilised in simulations.  The
calculations were performed on the HEALPix grid with $\Nside=512$. The
corresponding angular resolution of the grid ($7'$) is much higher than
the angular resolution of any of the existing or forthcoming cosmic ray
experiments but it allowed us to obtain an accurate sampling of arrival
directions of from-source \uhecrs.

Having these tools, it is straightforward to produce a map of arrival
directions of~$N$ \uhecrs, $\Nsrc$ of which come from a particular
source.  The procedure is as follows.  One takes the propagated spectrum
calculated with TransportCR for a source located at a given distance
from the Galaxy and samples it~$\Nsrc$ times, each time extracting some
nuclei with an energy~$E$ and charge~$Z$.  An observed arrival direction
of a cosmic ray is found then for each $(E,Z)$ pair (or, equivalently,
for each rigidity) using the mapping obtained with CRPropa by
backpropagation.  Finally, the remaining $N-\Nsrc$ events are generated
following the isotropic distribution. The whole process is repeated
multiple times in order to generate a large number of maps for each
source.

After that, we prepared maps of the relative intensity of the CR flux
and calculated their angular power spectra looking for the minimal
fraction~$\eta$ of from-source events in the whole sample allowing one
to reject the null hypothesis of an isotropic distribution
at a high confidence level.
The hypothesis of isotropy was tested using the following estimator:
\begin{equation}
	D(\text{sample}) = \frac{1}{\lmax}
		\sum_{\ell=1}^{\lmax}
		\frac{C_{\ell,\mathrm{sample}}
			-\langle C_{\ell,\mathrm{iso}}\rangle}
			{\sigma_{\ell,\mathrm{iso}}},
	\label{eq:D}
\end{equation}
where ``sample'' is either ``mix'' when applied to samples that 
contain a contribution from an UHECR source,
or ``iso'' when applied to isotropic samples, and
\Cl{} are coefficients of the angular power spectrum:
\begin{equation}
    \Cl = \frac1{2\ell+1}\sum_{m=-\ell}^{+\ell}|a_{\ell m}|^2.
	\label{eq:Cl}
\end{equation}
Thus, variables $C_{\ell,\mathrm{sample}}$, $\langle
C_{\ell,\mathrm{iso}}\rangle$ and $\sigma_{\ell,\mathrm{iso}}$ in
eq.~(\ref{eq:D}) are respectively the~\Cl{} observed in the sample
(either ``mix'' or ``iso''), the average and the standard deviation
of~\Cl{} for isotropic expectations, all of them calculated at a given
scale~$\ell$.
Coefficients $a_{\ell m}$ in eq.~(\ref{eq:Cl}) are the multipolar
moments of the spherical harmonics used to decompose
the relative intensity of the flux.

Since both~\Diso\ and~\Dmix\ in eq.~(\ref{eq:D}) are random variables,
one needs to compare their distributions.  It was assumed as the null
hypothesis~$H_0$ that arrival directions of a mixed sample of \uhecrs{}
obey an isotropic distribution.  We adopted the value of the error of
the second kind (the probability not to reject the null hypothesis when
it is false) $\beta=0.05$ and searched for a minimal fraction~$\eta$ of
from-source events in the total flux such that the error of the first
kind (the probability to reject the true null hypothesis)
$\alpha\lesssim0.01$.  For the sake of uniformity, all calculations of
the estimator~$D$ in Paper~1 were done with $\lmax=16$ though it was
remarked that the choice is not necessarily optimal, and slightly better
results can be obtained by adjusting~\lmax{} for each particular source,
depending on its angular power spectrum.

We used samples of sizes $N=100, 200,\dots, 500$ to cover the whole
possible range of \uhecrs{} to be detected by \keuso{} above
57~EeV~\cite{k-euso-ICRC2017-science}.  It was demonstrated that an
observation of $\gtrsim200\text{--}300$ events over the celestial sphere
(depending on the particular source) will allow testing~$H_0$ with the
above demands on~$\alpha$ and~$\beta$ providing the from-source events
form 10--15\% of the total sample.  In other words, registering a sample
of that size will allow one to detect a large-scale anisotropy arising
due to a nearby source with a high confidence level.\footnote{While the
	article was focused on \keuso, its results are valid for any other
	detector with a uniform exposure of the celestial sphere.}

Adjusting~\lmax{} to minimize~$\eta$ allows one to decrease numbers
obtained with the APS method by a few percent (from 2--5\% for $N=100$ to
1--2\% for $N=500$) most notably for Cen~A and Fornax~A, for which a
deviation of the lower multipoles from the isotropic distribution is
most pronounced, see Paper~1.
Obviously, this does not present a noticeable improvement of the result.

We remark the KKOS model provides a heavy mass composition of \uhecrs{}
at energies above 57~EeV thus resulting in much more fuzzy patterns of
arrival directions if compared with the case of a light (mostly proton and
helium) composition.
In this sense, our results are conservative since having more compact
patterns will allow obtaining less restrictive demands on the
minimal number of from-source events needed to reject the isotropy
hypothesis.

%_______________________________________________________________________
\section{Deep convolutional classifier}
\label{sec:cl_cnn}

As we have already mentioned above, a test statistics based solely on the
angular power spectrum cannot benefit from information about a pattern
of arrival directions of \uhecrs{} coming from a particular source. An
obvious way to overcome this limitation is to use some function of the expected and observed arrival direction maps. 
In what follows, we use a classifier based on the convolutional neural network architecture
as a test statistics. Recall that CNNs present a widely used subclass
of feed-forward neural networks designed specifically for pattern
recognition and image classification tasks~\cite{CNN}.

\begin{figure}[!t]
	\centering
	\figh{.5}{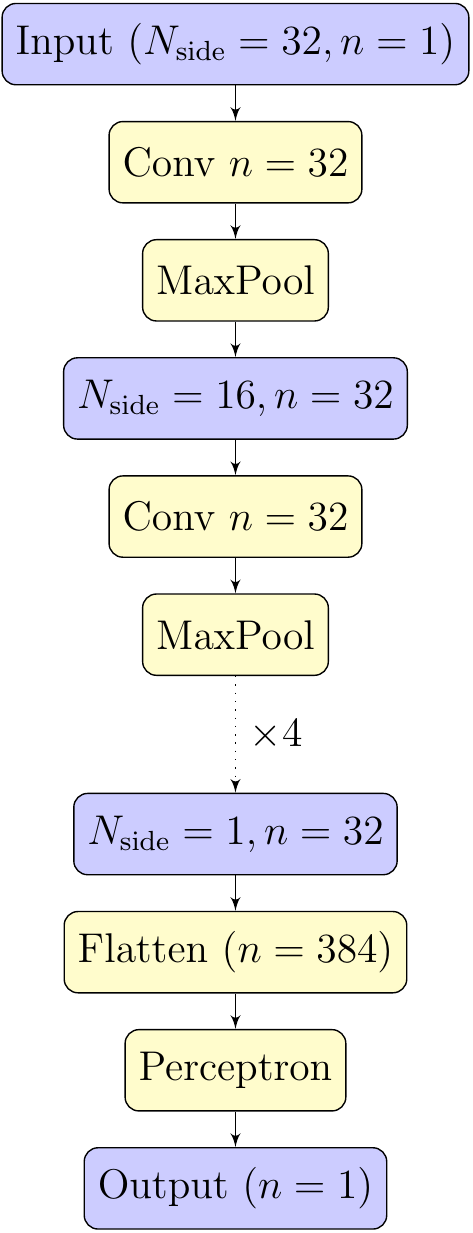}
	\caption{Architecture of the CNN developed in this work.
	Blue boxes are used to show feature vectors and maps. Yellow boxes
	show operations.}
	\label{fig:scheme}
\end{figure}

The basic idea behind CNNs is using local feature maps at different
scales to extract valuable information and perform a classification
task. A variety of CNN implementations exist for many programming
languages and platforms.  However, most of them support operations on
flat (two-dimensional) images only.  A number of implementations for
convolutional operations on the sphere were proposed
recently~\cite{SphericalCNN,Perraudin:2018rbt,Krachmalnicoff:2019zjh}.
We employed the publicly available code developed by
Krachmalnicoff and Tomasi~\cite{Krachmalnicoff:2019zjh}\footnote{%
	We applied two small patches to the original code,
	see~\cite{gitrepo} for details.},
which implements convolution and pooling (down-sampling) operations on
the HEALPix grid data with the help of Keras deep learning
library~\cite{keras}. The convolution operation on the HEALPix grid is
parameterized by 9 adjustable weights per feature map. The CNN
architecture developed in this work is shown in figure~\ref{fig:scheme}.

\begin{figure}[!t]
	\centering
	\figw{.48}{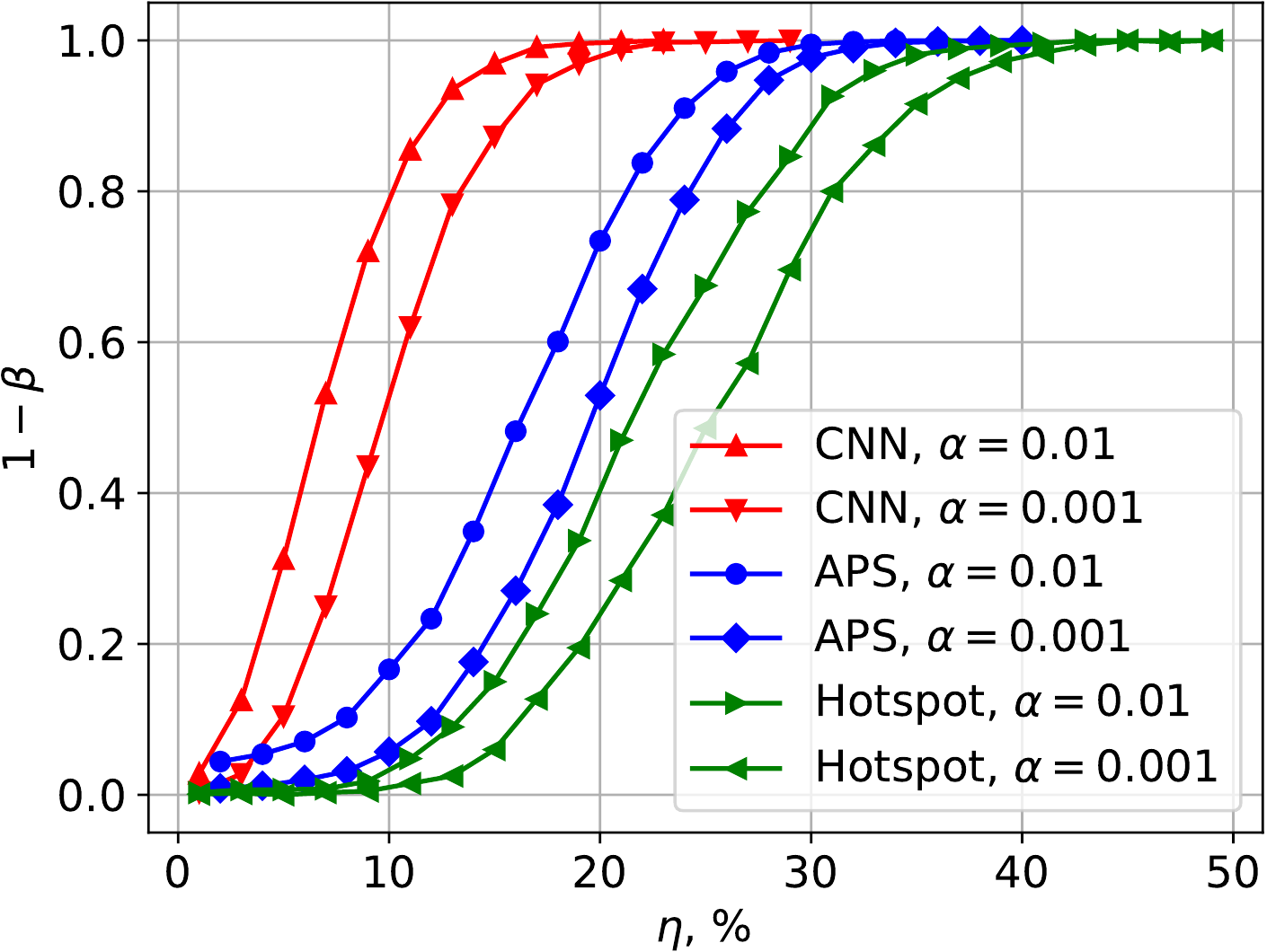}\quad
	\figw{.48}{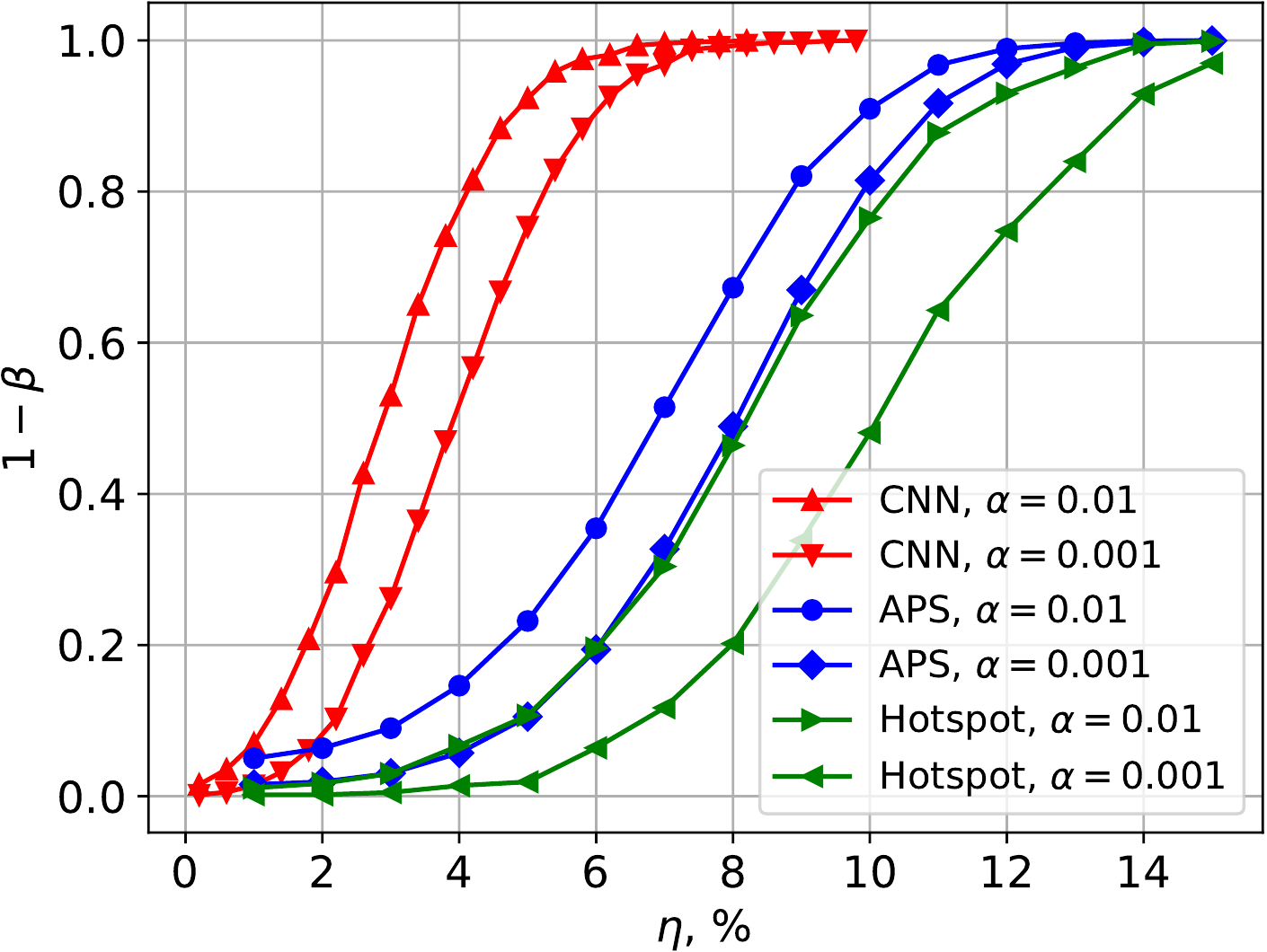}
	\caption{Dependence of the test power $1 - \beta$ on the fraction~$\eta$
		of UHECRs coming from M~87 in samples of size 100 (left) and 500 (right)
		for the approach based APS (blue) and with the
		suggested CNN (red) for $\alpha = 0.01, 0.001$.
		Green lines show the same for the approach based on a search
		for a ``hotspot'' at intermediate angular scales, see Remark~1.}

	\label{fig:test_power}
\end{figure}

The network takes one feature map ($n=1$ in figure~\ref{fig:scheme})
in the HEALPix grid with $\Nside=32$ as an input.\footnote{%
    With $\Nside=32$, the sphere is divided into 12,288 cells with the
    angular size of $1.83^\circ$, which is of the order of the
    angular resolution of UHECR experiments.}
Thirty-two feature maps
($n=32$) are built at the first step using the convolution operation
with $32\times9$ free parameters and max-pooling the image to
$\Nside=16$.
The sequence of convolutions and max-pooling operations is repeated
until reaching $\Nside=1$ with the persistent number of feature maps,
which means that each intermediate convolution operation has
$32\times32\times9$ trainable weights.  The rectified linear activation
function (ReLU) is used for all intermediate layers.  Finally, 32
feature maps with $\Nside=1$ are flattened and sent to a single-layer
sigmoid perceptron.  To avoid overfitting, we use an early-stop
technique.  Namely, we train our model for at most 1000 epochs and
interrupt training in case accuracy on validation data is not improving
for 10 epochs. Attempts to use other regularization techniques, such as
dropout and the L2 regularization demonstrated little benefit. 
The weights were optimized using the Adadelta adaptive learning rate
method~\cite{adadelta}. 
The output of the classifier, a number between~0 and~1 was used as the test
statistic. The minimal fractions~$\eta$ of from-source \uhecrs{} needed
to reject the null hypothesis of an isotropic flux with the same demands
on~$\alpha$ and~$\beta$ as above, are presented in
table~\ref{table:percentNN}.
Results obtained in Paper~1 are included for comparison (for $N\ge100$).
Figure~\ref{fig:test_power} provides two examples of the dependence of
the test power $1-\beta$ on the fraction of \uhecrs\ arriving from M~87
for both approaches, two sample sizes and $\alpha=0.01, 0.001$.
One can clearly see that the CNNs outperform the APS-based approach in
terms of the test power for any~$\eta$ and~$\alpha$.

\begin{table}[!t]

	\caption{Percentage of UHECRs arriving from the candidate sources
		in samples of sizes $N=50, 100,\dots,500$ such that the
		error of the first kind
		$\alpha\lesssim0.01$ for the null hypothesis of isotropy~$H_0$
		providing the second kind error $\beta=0.05$, obtained
		with the traditional approach based on the angular power
		spectrum (APS) and with the convolutional neural network (CNN)
		shown in figure~\ref{fig:scheme}.
		The result based on the APS was obtained for fixed $\lmax=16$.
		One source at a time was considered.}

	\label{table:percentNN}
	\medskip
	\centering
	\begin{tabular}{|l|c|c|c|c|c|c|c|}
		\hline
		Source    & Method &\,\,50\,\,&100&200&300&400&500\\
		\hline
		NGC 253   &APS& 24 &17 &  12 &  10  & 8 &   7 \\
					 &CNN& 12 & 7 & 4.5 & 3.67 & 3 & 2.6 \\ 
		\hline
		Cen A     &APS& 28 & 21 & 14 & 12  & 10 & 9   \\
					 &CNN& 16 & 11 & 7  & 5.67&  5 & 4.4 \\
		\hline
		M 82      &APS& 36 & 26 & 18 & 14 & 12   & 11  \\
					 &CNN& 20 & 12 &  7 & 6  & 4.75 & 4.2 \\
		\hline
		M 87      &APS& 38 & 29 & 20 & 16 & 14  & 12  \\
					 &CNN& 22 & 14 & 9  &  8 & 6.25& 5.2 \\  
		\hline
		Fornax A  &APS& 28 & 19 & 13 & 11 &  9 & 8   \\
					 &CNN& 16 &  9 &  6 &  5 & 4.5& 3.8 \\
    	\hline
	\end{tabular}
\end{table}

\paragraph{Remark~1.} One might expect that an anisotropy arising due to
an impact of a nearby source can be found more effectively with a
traditional method similar to the one used for finding the famous
``hotspot'' discovered by the Telescope Array
experiment~\cite{2014ApJ...790L..21A} and other intermediate scale
anisotropies found by other experiments.  To verify the conjecture, we
have written a dedicated code and performed searches for a hotspot
taking M~87 as a source for two ``limiting'' cases, $N=100$ and $N=500$
events, for definiteness.  Fraction of from-source events varied in
intervals 0--50\% and 0--15\% in the first and the second cases
respectively.  We have searched for an excess in circles with radii
from~20$^{\circ}$ to~60$^{\circ}$ in 10$^{\circ}$ steps.  The results
are shown in figure~\ref{fig:test_power} in green.  It can be
seen that the hotspot approach fared a bit worse than the one based on
the APS and was much less sensitive than the CNN. Thus, the approach is
not necessary the best in all possible cases though it can definitely be
very efficient in certain tasks.  A comprehensive investigation of
anisotropies at small and intermediate angular scales can be a subject
of a separate study.

\begin{figure}[!t]
	\centering
	\figw{.46}{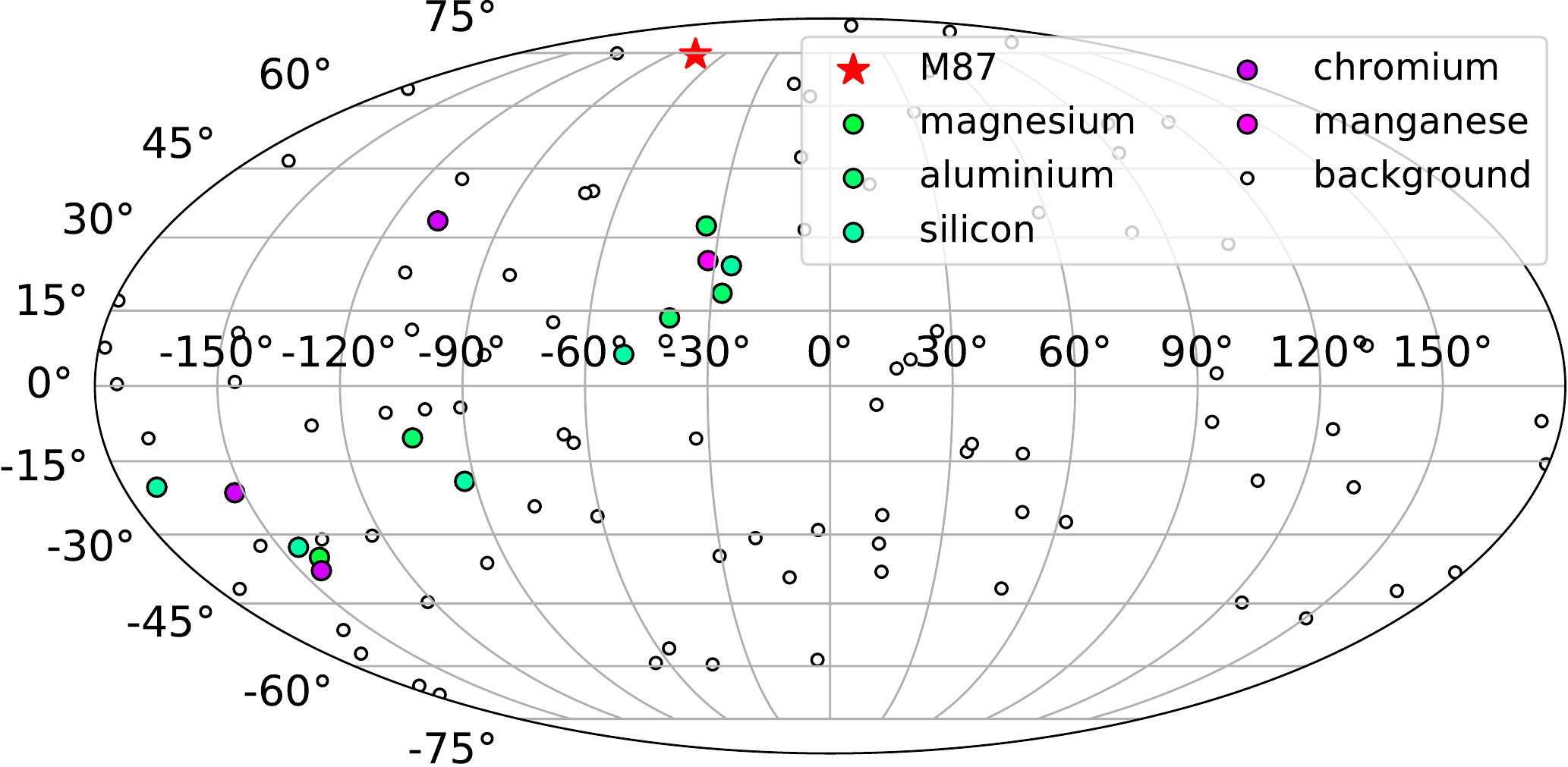}\quad \figw{.46}{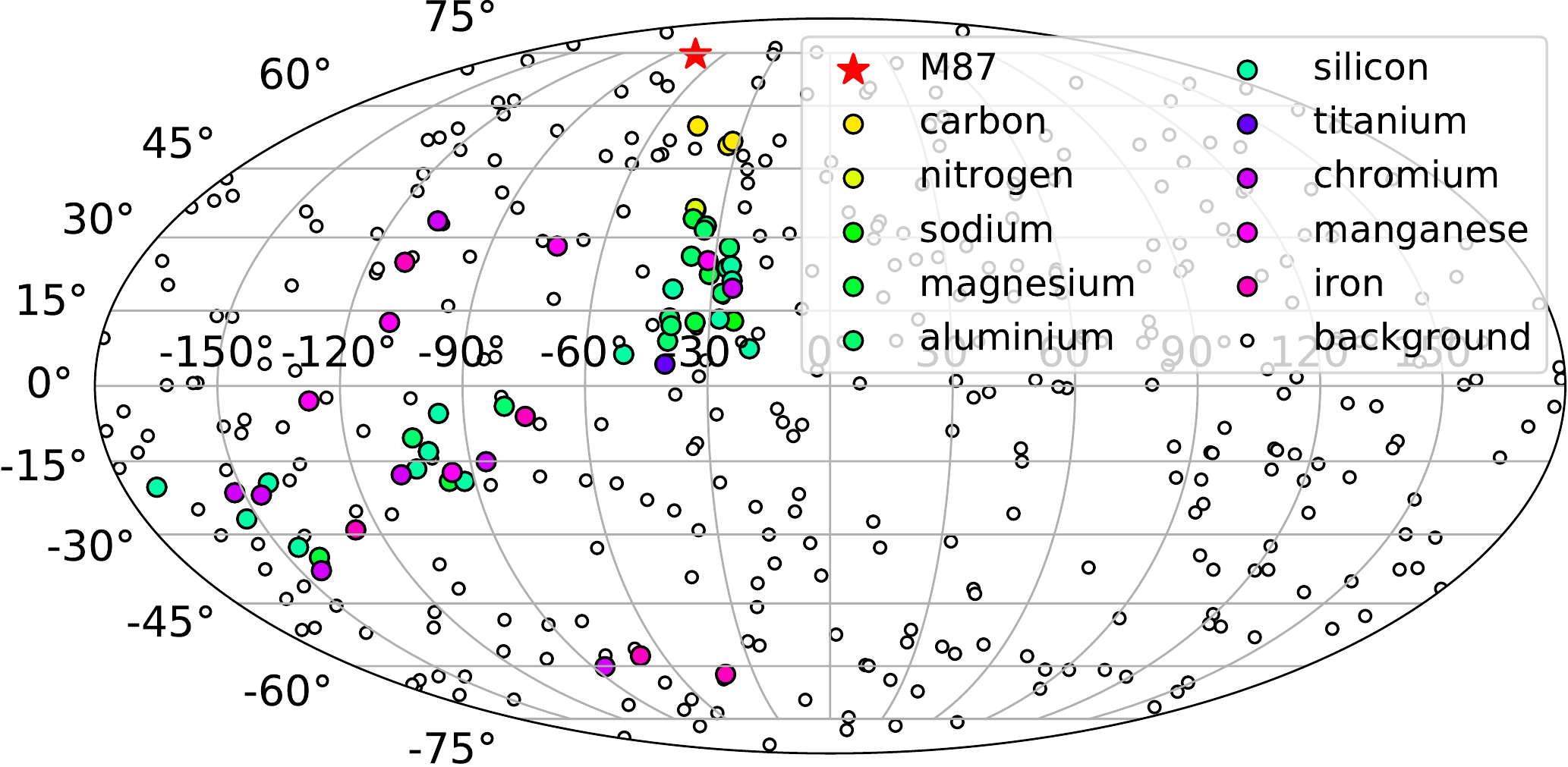}\\
	\caption{Left: an example of 14 \uhecrs{} with energies above
	57~EeV coming from M~87
	on the background of 86 events distributed isotropically
	(i.e., 14\% in the sample of 100).
	A sample like this allows testing~$H_0$ with the CNN.
	Right: an example of a sample with the same percentage
	of \uhecrs{} coming from M~87 (14\%) that allows testing~$H_0$
	in the classical approach based on the APS.
	Here, the total size of the sample equals 400, cf.\
	table~\ref{table:percentNN}.
	The maps are in the Mollweide projection with galactic coordinates.
	}
	\label{fig:M87}
\end{figure}

Notice the minimal fractions of from-source events needed to find an
anisotropy with CNNs decreased drastically in comparison with those
obtained with the classical approach.  Consider for example M~87, which
poses the most complicated case in comparison with the other sources due
to the fuzzy pattern of UHECR arrival directions, see the right panel of
figure~\ref{fig:M87} and plots in Paper~1.  In the classical approach,
one needed to register 400 events over the celestial sphere to
verify~$H_0$ for the given restrictions on~$\alpha$ and~$\beta$,
providing the flux of~\uhecrs\ arriving from M~87 comprises 14\% of the
total.  With the CNN, one needs to register only 100 events over the
sphere to test the hypothesis under the same conditions.  This means
even a pattern like the one shown in the left panel of
figure~\ref{fig:M87} with 14 \uhecrs{} coming from M~87 and 86
forming the isotropic background allows testing~$H_0$ by the CNN.

The situation is similar for the other sources.
Notice one needs to
register less than 100 events over the whole celestial sphere to verify the
null hypothesis providing any of the sources generates $\sim$10--15\% of
the total flux.
Alternatively, given a sample of a fixed size, one will be able to
test~$H_0$ with the CNN with a contribution from a source that it is at
least two times less than the one needed in the APS-based approach.

We also tried to use maps with cells of other sizes for the classifier
input but found that the minimal fraction of from-source events
necessary to test~$H_0$ only moderately depends on $\Nside$ for
$\Nside\geq32$ but grows substantially for maps with $\Nside\leq16$.

\paragraph{Remark~2.} It was considered in Paper~1 if an anisotropy
arising from a nearby source at energies above 57~EeV can be found by
the existing experiments at lower energies. It was demonstrated that a
source providing 10--15\% of the flux beyond 57~EeV can only contribute
1--1.5\% at energies above 8~EeV, at which a dipole anisotropy was found
by the Pierre Auger collaboration~\cite{Auger-dipole-2017}, and there is
very little difference between sources located at distances 3.5~Mpc and
20~Mpc in this respect.  The approach based on the angular power
spectrum and the estimator~$D$ defined in eq.~(\ref{eq:D}) did not allow
us to find an anisotropy from a nearby source even in case the total
sample contains 50,000 events, which is close to the size of the Pierre
Auger data set taking into account their field of view.  To the
contrary, the CNN presented above is able to distinguish a pattern of
from-source events comprising a much smaller fraction.  For example, it
is able to verify~$H_0$ with a pattern produced by \uhecrs{} with
energies above 8~EeV coming from Cen~A if they constitute as little as
$\sim1\%$ of
a sample of that size, with the same cuts on~$\alpha$ and~$\beta$.  This
suggests applying the technique to the analysis of data of the existing
experiments.

%______________________________________________________________________
\section{Model dependence. Towards a unified test statistic}

One of disadvantages of the neural network approach is the resulting
model opaqueness. Features used by the trained classifier are hard to
extract and interpret. By this reason, it is important to check the test
statistic resistance to variations of the physical model assumed during
training the neural network.  In this section, we present results of
this study and propose some ways to make the test statistic more
universal.  We also consider more realistic cases than the one
discussed above, namely, a situation when the source is not known
a~priori and when there are more than one nearby source contributing
to the UHECR flux and generating a kind of anisotropy.

\subsection{Galactic magnetic field model uncertainties}
\label{subsec:gmfs}

As it was mentioned in section \ref{sec:method}, the JF12 model of the
large-scale Galactic magnetic field~\cite{JF12,JF12b} was used for
training the neural networks presented above, as well as within the
traditional approach employed in Paper~1.  However, the GMF
is not known accurately yet, and some modifications of the JF12 model
have been suggested, as well as a number of alternative models.  We have
considered four recent models for the large-scale GMF, namely, the JF12
model as modified by the Planck Collaboration~\cite{Planck} (JF12P
below), a version of the JF12 model suggested by Kleimann
et~al.~\cite{Kleimann_2019} (JF12K), a model by Pshirkov
et~al.~\cite{PTKN} (PTKN) and one of the recent models suggested by
Terral and Ferri\`{e}re~\cite{TF17} (TF17).  We shall briefly recall the
main features of these models below but let us begin with the JF12
model since two of the other models are based on it.

The large-scale GMF in the JF12 model is described by three regular
components (a spiral disk field, a toroidal halo field and a poloidal
X-shaped field), a turbulent field and an extended random halo field. The
disk component of the turbulent field is modelled following the same
spiral structure as the regular component. The JF12 model has 36 free
parameters in total, which are constrained by radio observations of the
Faraday rotation of extra-galactic radio sources, measurements of the
polarized synchrotron emission of cosmic-ray electrons in the regular
component of the GMF and by measurements of the total synchrotron
intensity.

The Planck Collaboration modified the JF12 model
to match the experimental data of the Planck satellite on
polarized synchrotron emission at 30~GHz in conjunction with the
so-called z10LMPDE cosmic-ray lepton model~\cite{Planck}.  First, the
amplitude of the random component was changed to correct the degree of
ordering in the field near the Galactic plane.  The amplitude of the
X-shaped component was lowered, and the amplitude of the (otherwise
leading) coherent component of the Perseus arm was reduced.  The JF12
model was further modified to distribute the random component more
evenly through alternating spiral arm segments.  The simple generation
of the ordered random component in the JF12 model was kept intact by
scaling up the coherent component.  We have implemented the JF12P model
for CRPropa~3, and it is now available online.

Kleimann et~al.\ modified the JF12 model in two
aspects~\cite{Kleimann_2019}.  First, they inserted transitional layers
at the inner and outer rims of the spiral disk in which incoming and
outgoing magnetic field lines are redistributed, resulting in the spiral
field being fully divergence-free also at its inner and outer
boundaries.  As a result, a truly solenoidal large-scale spiral field
was obtained.  The second change relates to the poloidal X-shaped
component of the magnetic field in the JF12 model and serves to remove
the sharp kinks of field lines that are present in the original model at
the Galactic midplane.  These kinks were removed by either a numerical
convolution technique, or analytically replaced with smooth parabolic
inserts, which also fully satisfy the divergence constraint.

Pshirkov et~al.\ suggested two models of the GMF based on experimental
data on rotational measures of extra-galactic radio sources~\cite{PTKN}.
Both models include a halo and a disk component, with the latter coming
in two versions: an axisymmetric one (ASS), with the direction of the
magnetic field being the same in two different arms, and a bisymmetric
(BSS), with the direction being opposite.  The fitting procedure
done by Pshirkov et~al.\ could not discriminate between the ASS and
BSS models.  We employed the BSS model of magnetic fields in the disk
(together with the halo model), similar to the one used in the TF17
model.  Notice the PTKN model does not have a random component.

Finally, Terral and Ferri\`{e}re~\cite{TF17} considered as much as 35
models of the total (halo plus disk) magnetic field, each composed of
one of their seven anti-symmetric halo field models plus one of the five
symmetric disk-field models.  The work was based on analytical models of
spiralling, possibly X-shape magnetic fields developed in their earlier
paper~\cite{TF14} and on data on Faraday rotation measures of
extragalactic point sources. In the present work, we employed a so
called Ad1 bisymmetric disk field and C1 halo field as implemented in
CRPropa, which were found by Terral and Ferri\`{e}re to be among the
most favourable of the models.

\begin{figure}[!ht]
	\centering
	\figw{.46}{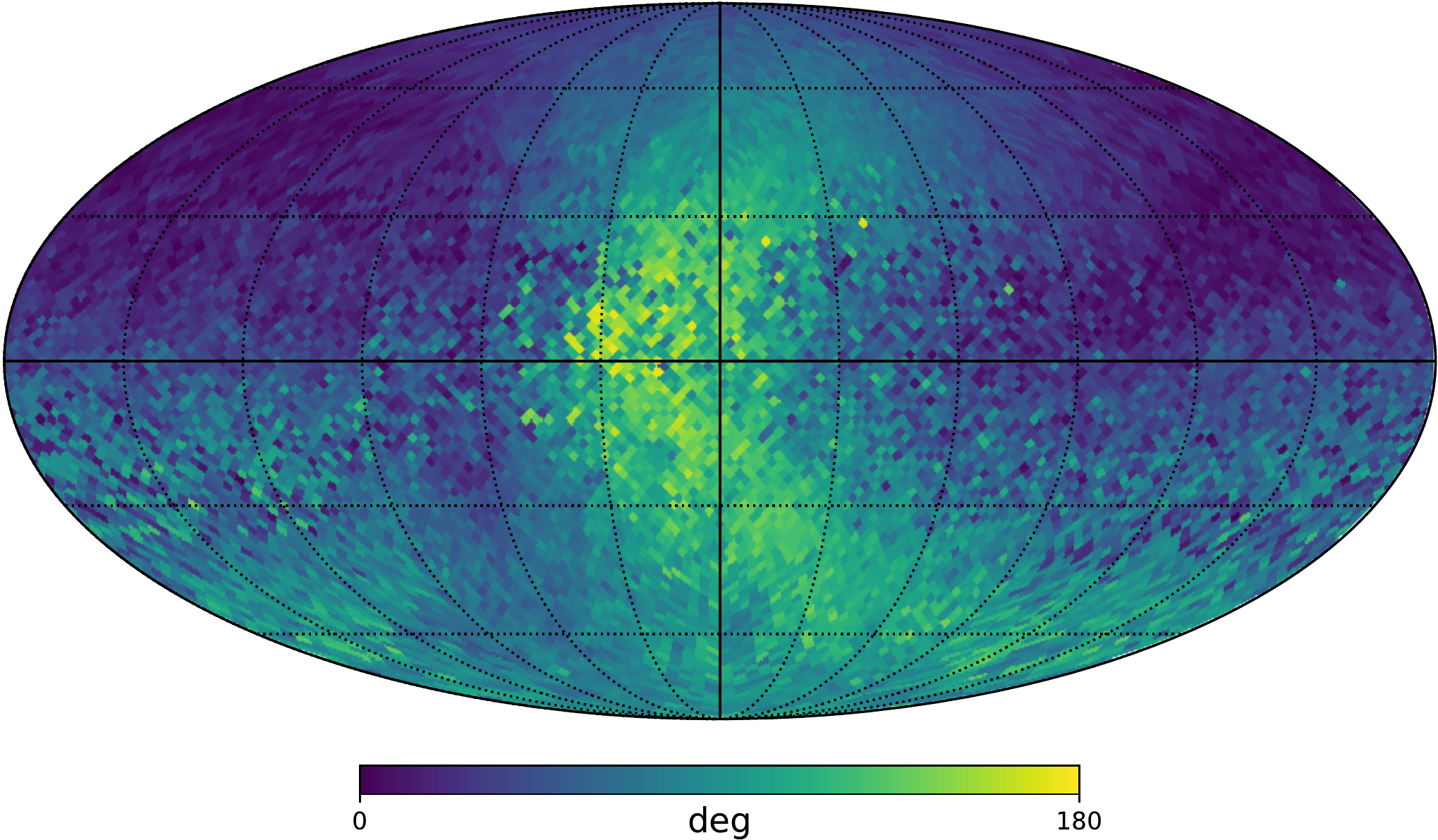}\\
	a)\figw{.46}{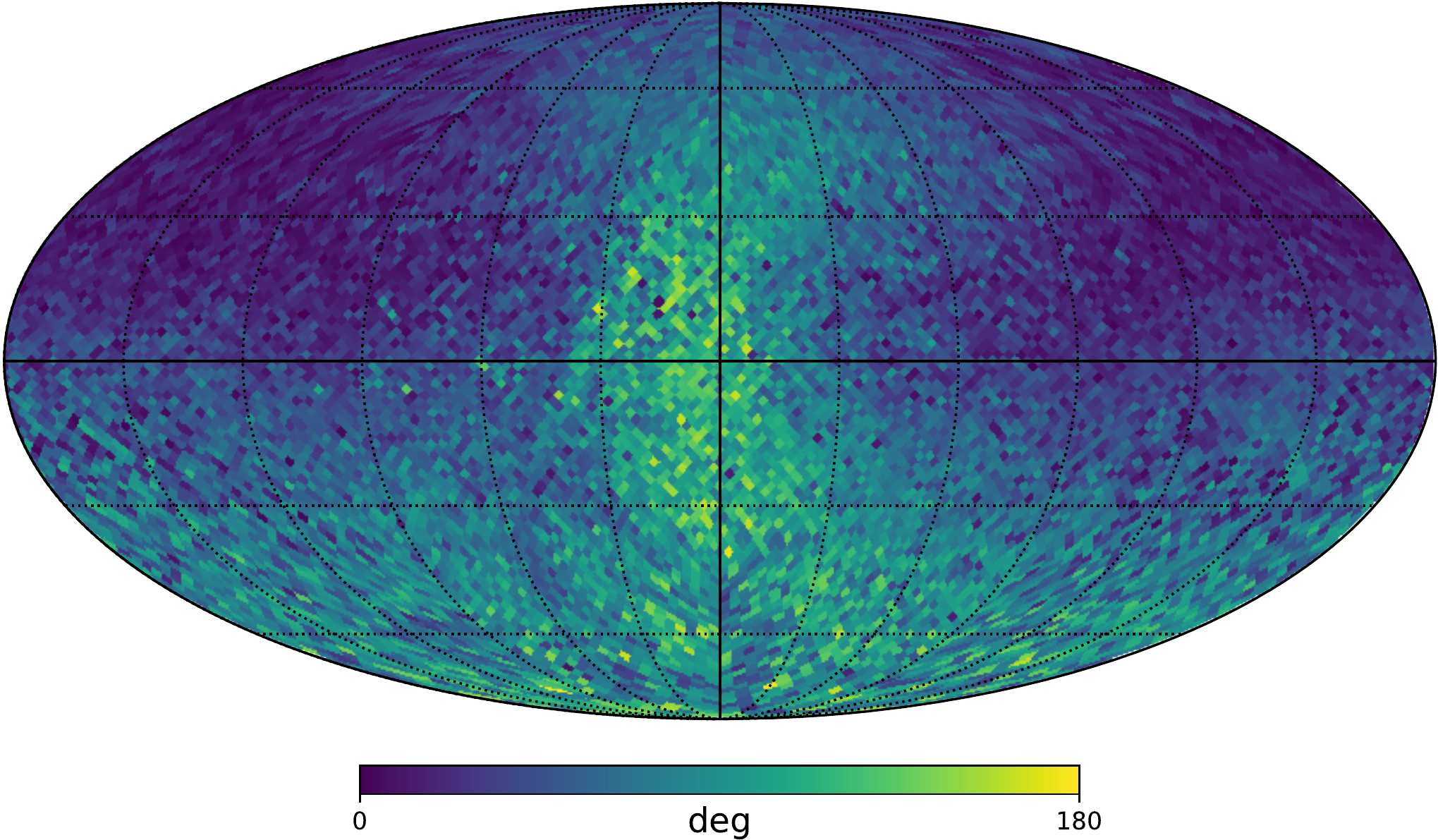}\quad b)\figw{.46}{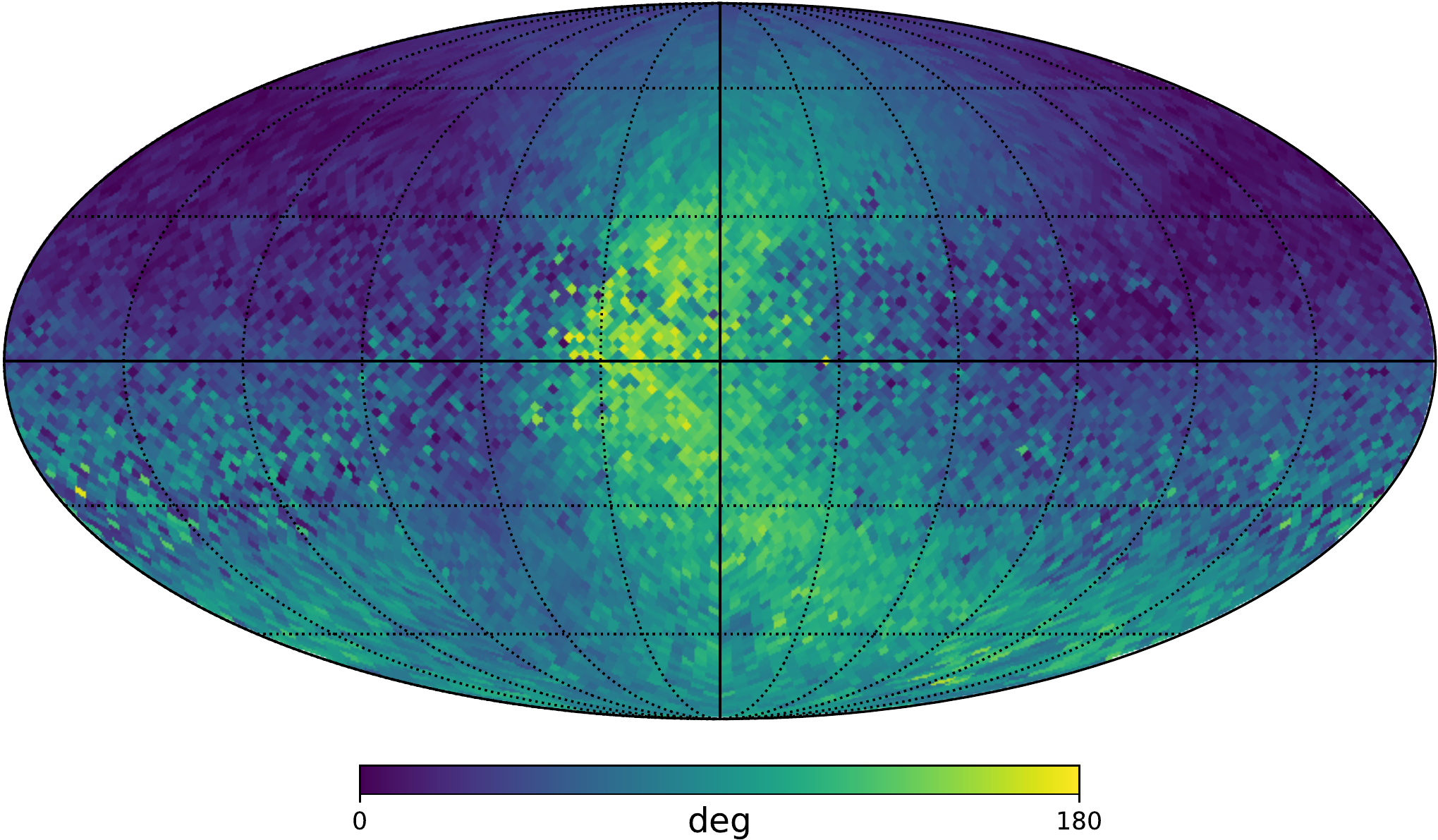}\\
	c)\figw{.46}{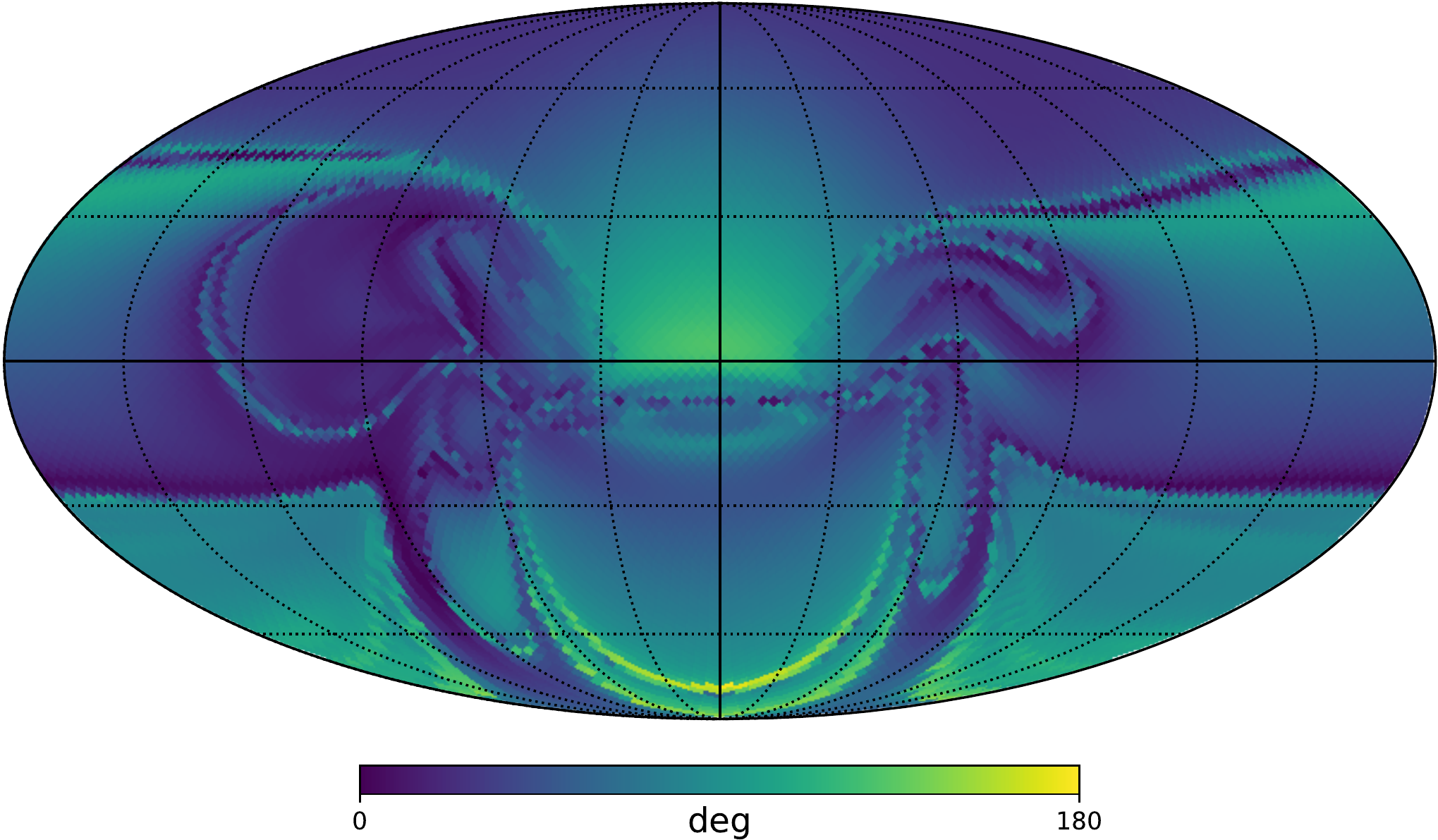}\quad d)\figw{.46}{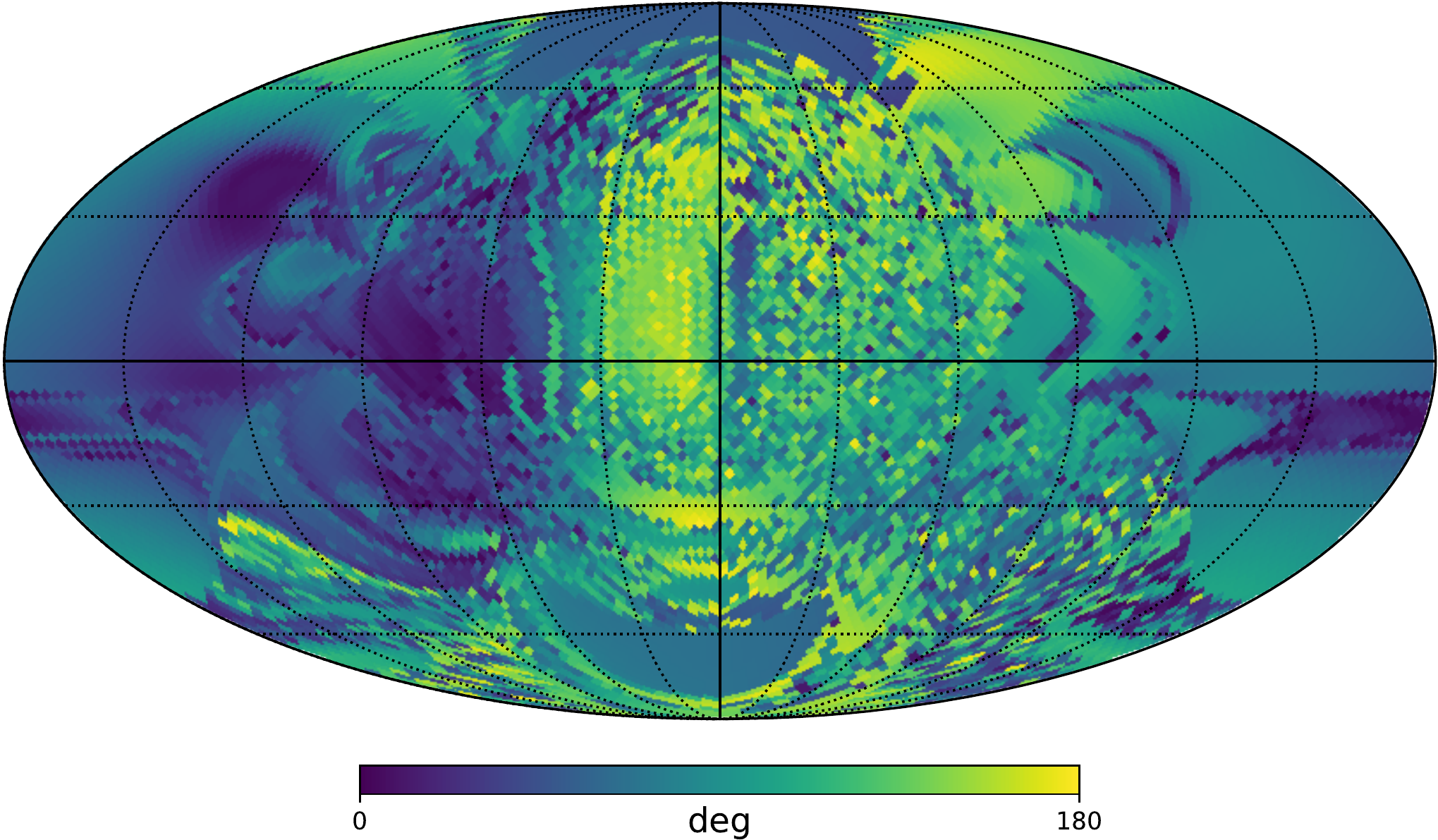}
	\caption{Absolute values of deflections of a 100~EeV iron nuclei
	from an initial arrival direction on its way to the Solar system in
	different models of the large-scale Galactic magnetic fields.
	Top panel: the JF12 model.
	Other panels: a)~JF12P, b)~JF12K, c)~PTKN d)~TF17.
	Galactic coordinates in the Mollweide projection are used.}
	\label{fig:GMF_models}
\end{figure}

\begin{figure}[!ht]
	\centering
	a)\figw{.46}{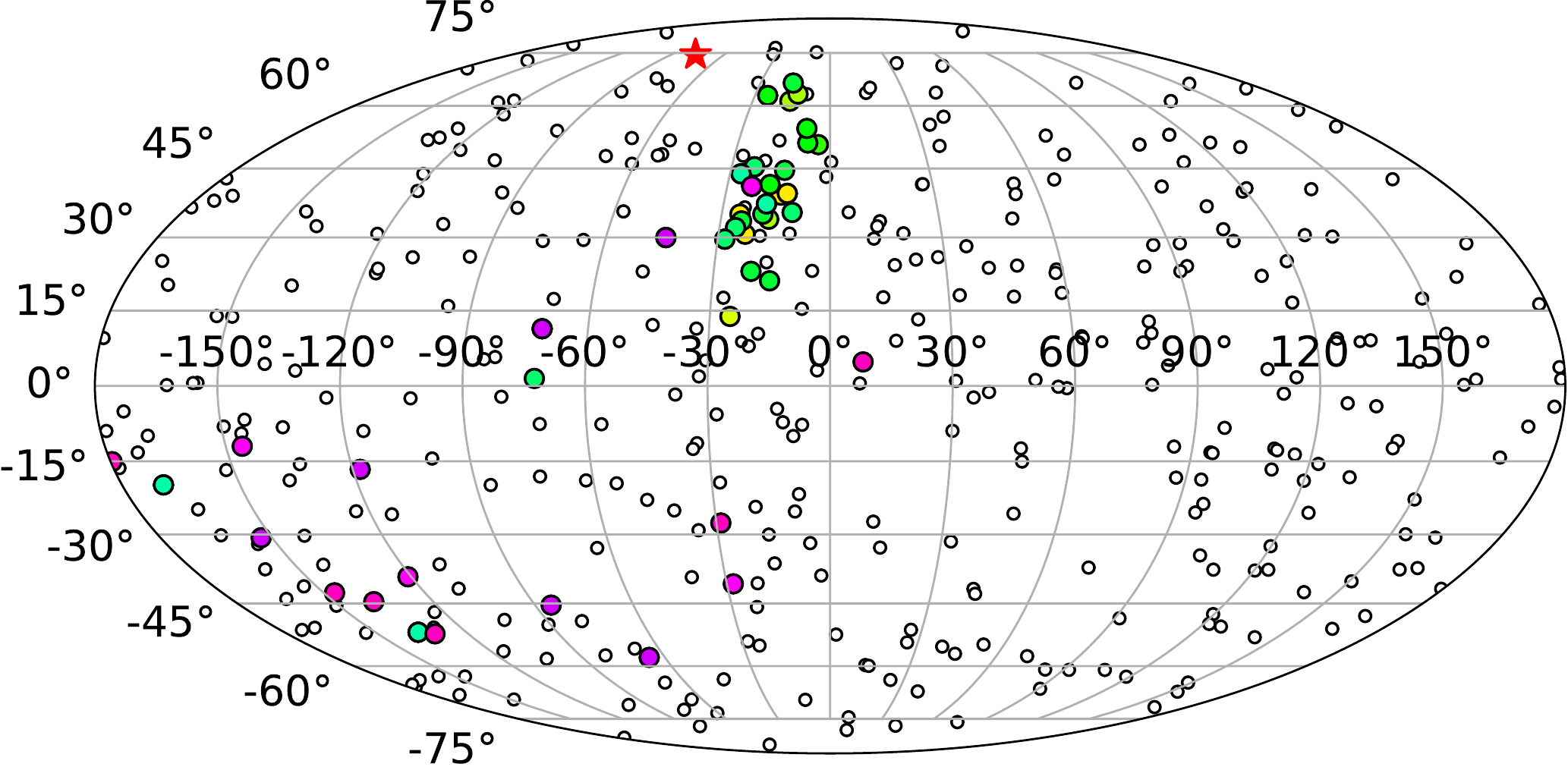}\quad b)\figw{.46}{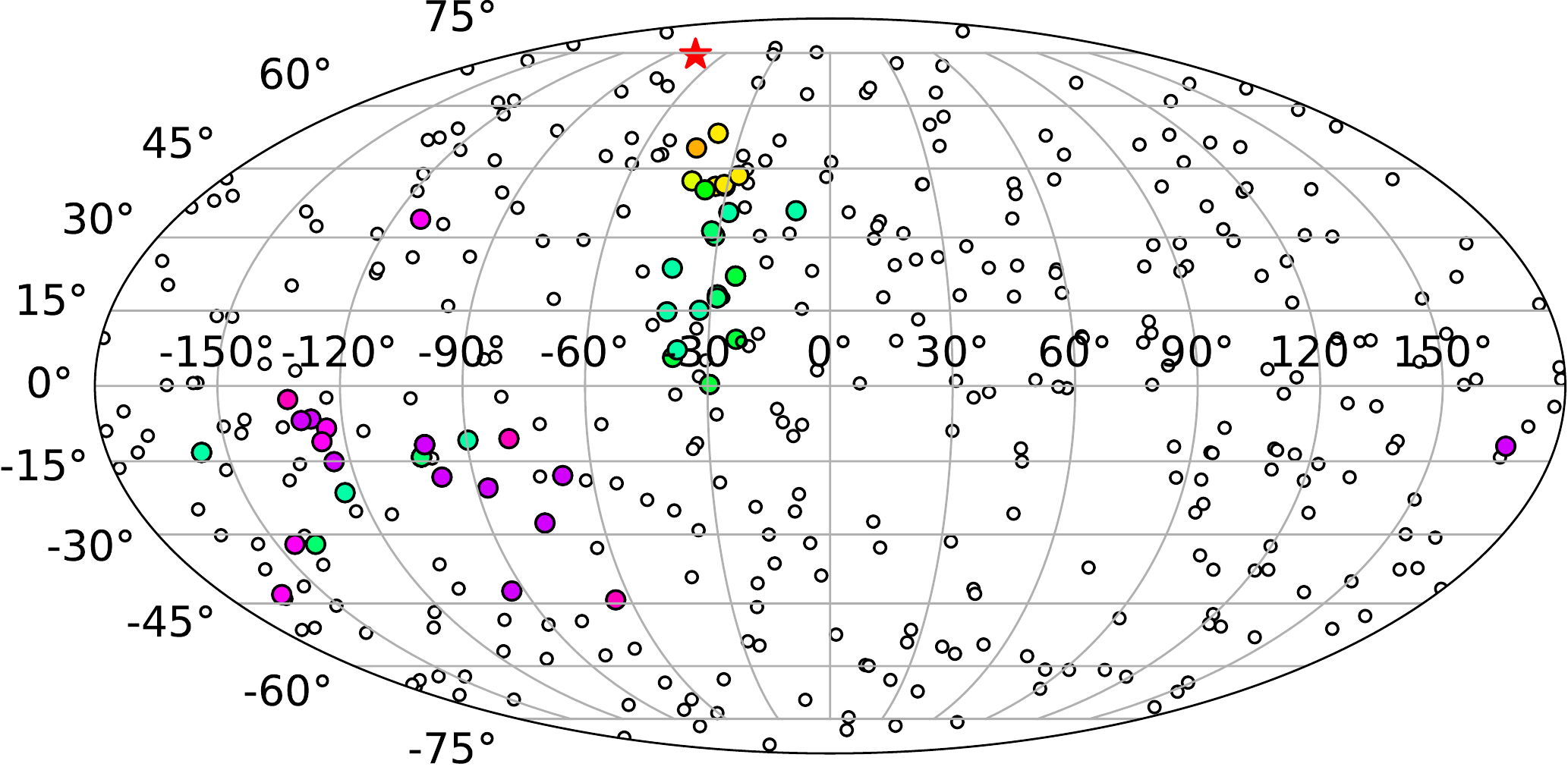}\\
	c)\figw{.46}{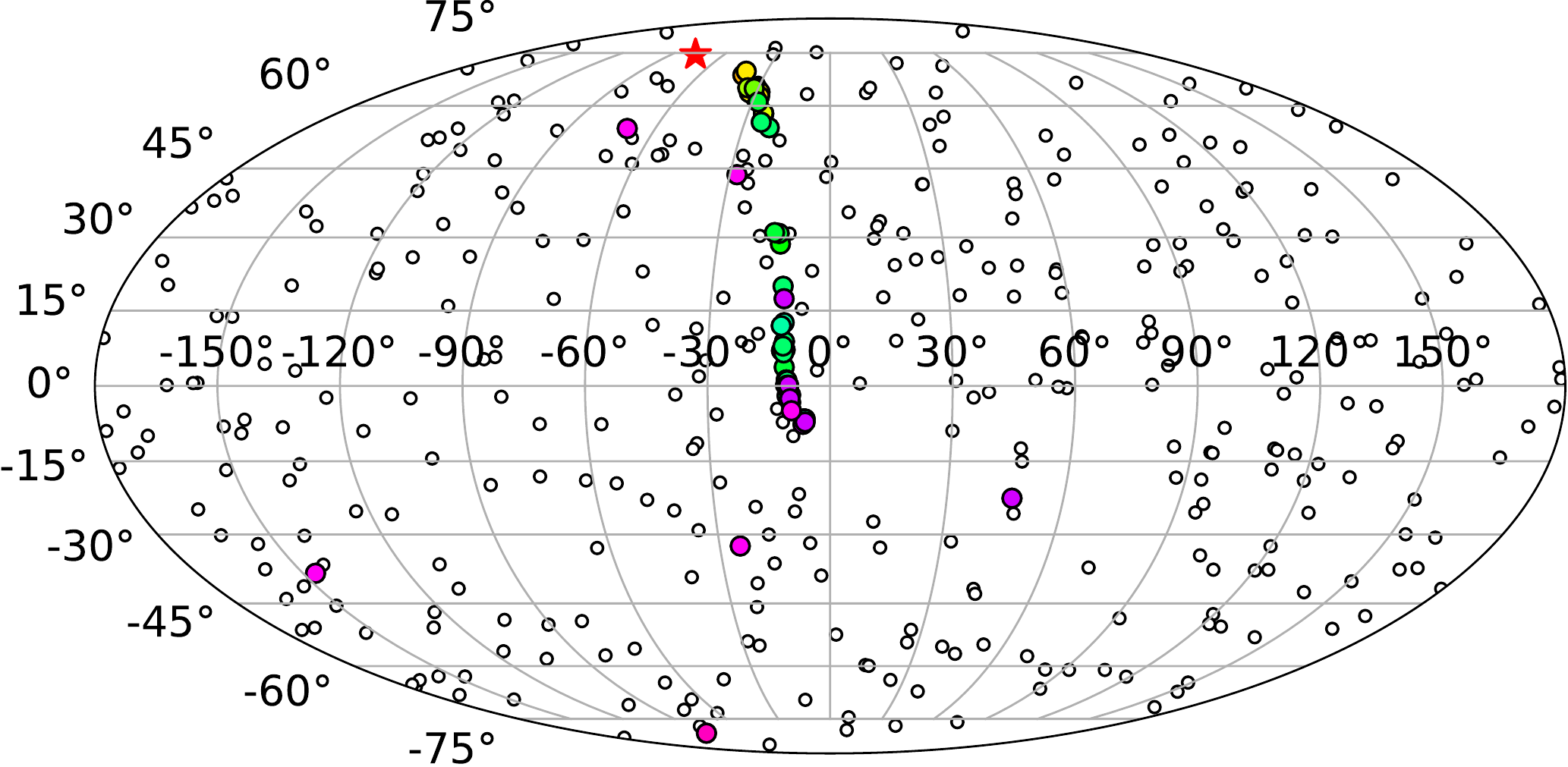}\quad d)\figw{.46}{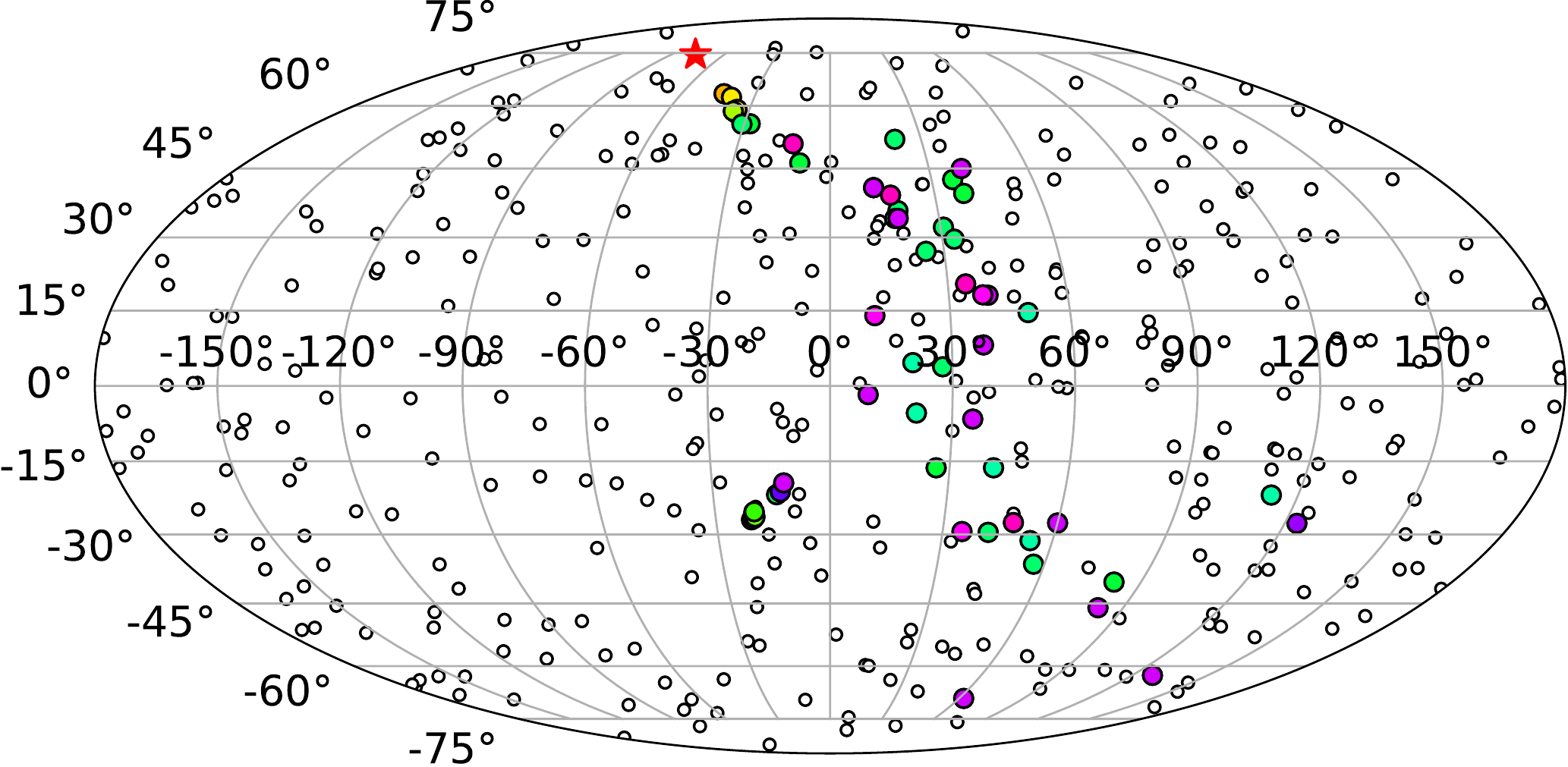}
	\caption{Examples of samples of 400 \uhecrs{} with energies above
	57~EeV with 14\% of them coming from M~87 obtained with different models
	of the GMF: a)~JF12P, b)~JF12K, c)~PTKN d)~TF17. Compare them with
	the right panel in figure~\ref{fig:M87} showing the same for JF12.
	}
	\label{fig:M87_vs_GMF}
\end{figure}

Figure~\ref{fig:GMF_models} illustrates how an UHECR is deflected from
its initial arrival direction on its way to the Solar system in
different models of Galactic magnetic fields.  In its turn,
figure~\ref{fig:M87_vs_GMF} gives examples of how these four models act
on \uhecrs{} arriving from a particular source.  Compare these patterns
of arrival directions with those shown in the right panel of
figure~\ref{fig:M87}.  It is clearly seen the TF17 model results in a
pattern that deviates from the one in the JF12 model most of all.

% ______________________________________________________________________

To examine the robustness of the test statistic introduced in
section~\ref{sec:cl_cnn}, we have tested the original neural network
classifier trained on the JF12 model on event sets generated assuming
the four alternative GMF models described above.  To make the test
statistic more universal, we modified the early-stop procedure to avoid
overfitting when training the classifier.  Namely, we used a set of
events prepared assuming the PTKN model of the GMF as validation data.
Results of the tests are presented in table~\ref{table:B_dependence}.%
\footnote{In a few cases, the classifier trained on maps with the number
of events different from the one given in columns gave better results.
We present results obtained with the best classifier in this case.}
Notice that similar to the results presented above, CNNs were trained
separately for each particular source and the number of incoming
\uhecrs{}, with the background flux being isotropic.

\begin{table}[!ht]
 	\caption{Results of tests of the CNNs trained with the JF12 model of the 
	GMF versus data sets generated assuming other models.
	Here, JF12 is the original
	model~\cite{JF12,JF12b}, JF12P is the JF12 model modified by the Planck
	Collaboration~\cite{Planck}, JF12K is a version of the JF12 model
	suggested by Kleimann et~al.~\cite{Kleimann_2019},
	PTKN is a model by Pshirkov et~al.~\cite{PTKN} and
	TF17 is a model by Terral and Ferri\`{e}re~\cite{TF17}.
	Numbers mean the same percentage as defined in table~\ref{table:percentNN}.
	A dash means a solution was not found.
	See figure~\ref{fig:tables} below for an illustration of the data for M~87.
 	}
 	\label{table:B_dependence}
 	\medskip
 	\centering
 	\begin{tabular}{|l|c|c|c|c|c|c|c|}
 		\hline
 		Source     & GMF   & \,50\, & 100    & 200    & 300    & 400    & 500 \\
 		\hline       
					  &  JF12 & 14 & 10 & 6   & 5   & 3.75& 3.6 \\
					  & JF12P & 20 & 13 & 8   & 6.7 & 5.5 & 5.2 \\
		NGC 253    & JF12K & 14 &  9 & 5.5 & 4.3 & 3.75& 3.4 \\
					  &  PTKN & 28 & 16 & 9.5 & 8   & 6.5 & 5.8 \\
					  &  TF17 & 64 & 42 & 24  & 18.3&14.25& 12  \\ 
		\hline       
					  & JF12  & 16 & 12 & 7.5 & 6   & 5.25& 4.8 \\
                 & JF12P & 22 & 14 & 10  & 8   & 6.75& 5.8 \\
		Cen A      & JF12K & 16 & 11 &  8  & 6.3 & 5.5 & 4.8 \\
					  & PTKN  & 20 & 12 &  8  & 6   & 5.25& 4   \\
					  & TF17  & 40 & 25 & 19  & 13.7& 12  & 10.4\\
		\hline       
					  & JF12  & 20 & 14 & 8.5& 6.3 & 5.5 & 4.6 \\
					  & JF12P & 26 & 17 & 10 &  8  & 6.5 & 6.2 \\
		M 82       & JF12K & 24 & 15 &  9 & 7.3 & 6   & 5   \\
					  & PTKN  & 20 & 12 &  7 & 5.3 & 4.25& 3.8 \\
					  & TF17  & 30 & 20 &12.5& 9.7 & 7.75& 6.2 \\
		\hline       
					  & JF12  & 24 & 17 & 11 & 8.3 & 7.5 & 6.6 \\ 
					  & JF12P & 46 & 22 &17.5&11.67& 10&8.6 \\  
		M 87       & JF12K & 26 & 17 &11.5& 9   & 7.75& 7   \\
					  & PTKN  & 34 & 18 & 11 & 8.7 & 7   & 5.6 \\
					  & TF17  & -- & 79 &44.5&33.7 &26.5&24.2 \\
		\hline       
					  & JF12  & 16 & 11 & 7.5 &  6 & 4.75& 4.6 \\ 
					  & JF12P & 26 & 17 &10.5 & 8.3& 7   & 6.6 \\
		Fornax A   & JF12K & 16 & 10 & 6.5 & 5  & 4.5 & 3.8 \\
					  & PTKN  & 26 & 16 & 10  &7.7 & 5.5 & 5.4 \\
					  & TF17  & 58 & 38 & 18  & 15 &10.75& 9.8 \\
		\hline
	\end{tabular}
\end{table}

In general, the CNNs trained with the JF12 model demonstrate excellent
robustness against the JF12K model, for which they sometimes produce
even better results, possibly due to more compact patterns of arrival
directions.  The JF12P model is more complicated for the CNNs but
results are still good for almost all sources except M~87, for which the
minimum required fraction of from-source events grows strongly
for small samples.
The PTKN model acts sufficiently well sometimes
demonstrating results that are even better than for the original JF12
model (see the case of M~82).  Finally, the TF17 model presents the
biggest problem for the CNNs trained for JF12. Still, except for M~87,
even this model of the GMF works well enough to allow detecting a nearby
source providing the size of the total sample is big enough.
Thus, one can conclude that the suggested CNNs demonstrate good
efficiency providing the main features of the Galactic magnetic field
are understood sufficiently well.  Notice the new CNNs perform slightly
worse for the JF12 model than those used in table~\ref{table:percentNN}
due to the early-stop procedure, which is non-optimal for them.  An
important point is the CNNs can be easily updated as soon as our
knowledge of the large-scale GMF improves.

%_______________________________________________________________________
\subsection{Testing the CNNs against different mass compositions}
\subsubsection{Mass composition as a function of the propagation length
in the KKOS model}

\begin{figure}[!t]
	\centering
	\figw{.6}{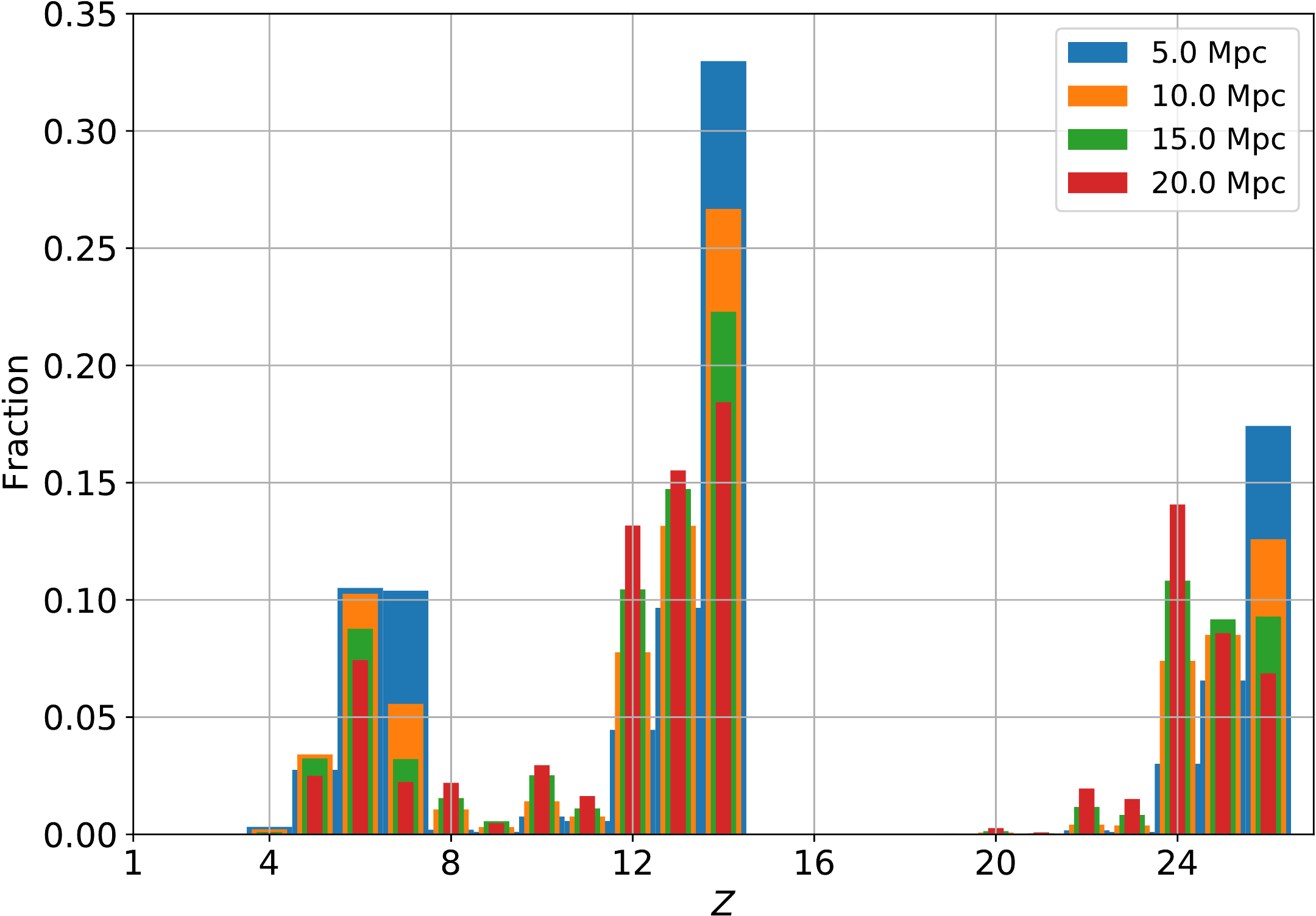}
	\caption{Dependence of the mass composition of \uhecrs{} above
	57~EeV on the distance to the source in the KKOS model.
	The height of bins is normalised to the total flux.
	$Z$ is the atomic number of a nuclei.}
	\label{fig:mass_vs_distance}
\end{figure}

It is known that mass composition of \uhecrs{} registered on Earth
depends on the distance to the source.  One and the same AGN located at
different distances from Earth will generate \uhecrs{} with possibly
distinguishable mass compositions due to interactions nuclei face on
their way from the source.  This results in different patterns of their
arrival directions.  Figure~\ref{fig:mass_vs_distance} illustrates how
mass composition of \uhecrs{} registered at Earth depends on the
propagation length in the KKOS model.

We used distances given in the catalogue~\cite{vanvelzen} rounded to
0.5~Mpc for the sources considered above.  On the other hand, the
NASA/IPAC Extragalactic Database (NED) provides a whole range of
distances to the objects.  We took the mean values plus/minus standard
deviations and tested the CNNs used to obtain
table~\ref{table:percentNN} against maps of arrival directions obtained for
samples arriving from the minimum and maximum distances defined this
way.  It was found that percentage of from-source events necessary to
verify the original null hypothesis~$H_0$ deviates by at most~10\% from
the numbers given in table~\ref{table:percentNN} even for Fornax~A, which
has the biggest uncertainty in the distance, thus demonstrating an
excellent robustness of the CNNs in these tests.

%_______________________________________________________________________
\subsubsection{Comparison with other mass compositions}

Mass composition of UHECRs is a subject of ongoing studies and debates.
It is known that results of the Telescope Array (TA) experiment witness
in favour of a light mass composition in the whole energy range above a
few EeV~\cite{2019ICRC...36..280H} while the Pierre Auger Observatory
argues that the UHECR flux at around 2~EeV mostly consists of light
nuclei but becomes heavier at higher
energies~\cite{2019ICRC...36..482Y}.  The TA collaboration has recently
suggested a mass composition that fits well the data of both
experiments. It consists of protons (57\%), nuclei of helium (18\%),
nitrogen (17\%) and iron (8\%)~\cite{2019ICRC...36..280H}.  We will
refer to this composition as TA4.  We have considered how the CNNs
presented in subsection~\ref{subsec:gmfs} (trained with the KKOS mass
composition and the JF12 model of the GMF model) perform with this new
composition.  Besides this, we tested the CNNs against two other mass
compositions, one of which being in average two times lighter than the
original composition in the KKOS model, and another one being two times
heavier.  Compositions in these two models are probably unnatural.  We
introduced them just to evaluate the robustness of the test statistics
obtained in subsection~\ref{subsec:gmfs}.  Results of the tests are
presented in table~\ref{table:cnn_vs_mass}.

\begin{table}[!ht]
 	\caption{Results of tests of the CNNs used in
	subsection~\ref{subsec:gmfs} versus different mass compositions of
	\uhecrs. Here, ``Mass'' denoted as ``0.5'' corresponds to the mass
	composition that is in average two times lighter than that in the
	original KKOS model, ``2.0'' denotes a composition that is two times
	heavier. ``TA4'' denotes the 4-component composition
	suggested by the Telescope Array
	collaboration~\cite{2019ICRC...36..280H}.
	Numbers denote the same minimum percentage of \uhecrs{}
	necessary to verify the null hypothesis~$H_0$ as in all
	tables above.
	See figure~\ref{fig:tables} below for an illustration of the data for M~87.
	}
 	\label{table:cnn_vs_mass}
 	\medskip
 	\centering
 	\begin{tabular}{|l|c|c|c|c|c|c|c|}
 		\hline
 		Source   &Mass&\,50\,& 100 & 200 & 300 & 400 & 500 \\
 		\hline    
					&0.5 & 12  &    8 & 5.5 & 4   & 3.5 & 3.4 \\
		NGC 253  &2.0 & 18  &   13 &   9 & 6   & 5.25& 4.8 \\ 
					&TA4 &  26 &  15  & 10 & 7.33 & 6.25  & 5.4\\
		\hline    
               &0.5 &  14 &   9  &  6   & 4.67  & 4.25& 3.6 \\
		Cen A    &2.0 &  22 &  14  & 10   & 8     & 6.5 & 6   \\ 
               &TA4 &  14 &   8  &  4.5 & 3.3   & 2.75&  2.2 \\
		\hline    
               &0.5 & 10  &    6 & 3.5  & 2.3  &  2   & 1.6 \\
		M 82     &2.0 & 52  &   34 &  25  & 20   & 16.75& 14.8\\
					&TA4 &  8  &   5  &  3   &  2   & 1.75 & 1.4 \\
		\hline    
					&0.5 & 14  &  10  &  5.5 & 4.3  & 3.75 & 3  \\
		M 87     &2.0 & 60  &  35  &  27  & 22   & 17.5 & 16\\
					&TA4 & 24  &  14  &  8.5  & 6.67 & 5.5 &4.6\\
		\hline    
					&0.5 & 14  &  10  &   5   & 4.3 &  3.5 & 3 \\ 
		Fornax A &2.0 & 28  &  19  &  15   &10.67& 9.25 & 9 \\  
					&TA4 & 18  &  14  &  7    & 5.3 & 4    & 3.6 \\
		\hline
	\end{tabular}
\end{table}

As one could expect, the light composition (``0.5'') does not pose a
problem for the CNNs since patterns of UHECR arrival directions become
less fuzzy. More than this, the CNNs demonstrate even better results in
this case than with the original composition, c.f.\ results
corresponding to the JF12 model in table~\ref{table:B_dependence}. The
heavy (``2.0'') composition also worked fine for all sources except M~82
and M~87.
This does not come as a surprise since light nuclei
arriving from these two sources form long and narrow ``tails'' on the
celestial sphere while heavy ones are spread over huge regions, see the
right panel in figure~\ref{fig:M87} for M~87 and figure~4 in Paper~1 for
M~82.  As for the TA4 composition, results also depend on the particular
source.  The CNNs demonstrated good performance for Cen~A, Fornax~A and
especially M~82 but needed higher percentage of events coming from M~87
and NGC~253 to verify~$H_0$ for small samples.  Generally, we conclude
the CNNs exhibit good robustness to reasonable variations of the mass
composition used for their training.  Thus, the present uncertainty in
the composition of \uhecrs{} is not crucial for their application.

%_______________________________________________________________________
\section{Multiple nearby sources of UHECRs}

So far we considered the case of a single nearby source of \uhecrs{},
and the problem was to reject the null hypothesis~$H_0$ (isotropy)
providing an alternative hypothesis (anisotropy) is true.  To solve the
task, we trained a separate classifier for each particular source and
each size of the total sample to use it as a test statistic.  A possibly
more realistic situation is when there are two or more nearby sources
contributing to the flux of \uhecrs.  

In Paper~1, we estimated the average event number coming from the second
strongest source in case of identical sources to be roughly about~1/3 of
the main source contribution (see Table~3 in Paper~1).  Here, we have
evaluated the performance of the test statistics obtained for Cen~A
being the only nearby source in a situation when there is another strong
source, and test samples contain three times more events from Cen~A than
from that other source.  The tests were performed for the JF12 model as
well as for the other four models of the GMF.  The results are presented
in table~\ref{table:CenA_3to1}.  Notice that numbers given in the table
denote percentage of the \emph{joint} flux from Cen~A and the second
source relative to the total flux.

\begin{table}[!ht]
	\caption{%
	Application of the test statistics based on classifiers trained on
	Cen~A (rows 6-10 of table~\ref{table:B_dependence}) to samples
	containing an admixture of events from another source.  In test
	samples, the average number of events from the second source was
	three times less than from Cen~A.  Notation is the same as in
	table~\ref{table:B_dependence} except the percentage of the joint
	flux from Cen~A and the second source relative to the total flux is
	shown now.
	See figure~\ref{fig:tables} below for an illustration of the data for M~87.
	}
 	\label{table:CenA_3to1}
 	\medskip
 	\centering
 	\begin{tabular}{|l|c|c|c|c|c|c|c|}
 		\hline
		2nd source & GMF& 50 & 100 & 200 & 300 & 400 & 500 \\
 		\hline
					  &JF12 & 24 & 16 & 10.5& 8     & 7     & 6.8\\
					  &JF12P& 32 & 22 & 14  & 10.33 & 9.5   & 8.4\\
		NGC 253    &JF12K& 24 & 18 & 11.5& 8.67  & 7.25  & 6.8\\
					  &PTKN & 30 & 18 & 12  & 8.33  & 7.25  & 6.2\\
					  &TF17 & 60 & 40 & 27  & 18.67 & 17.25 & 15 \\
 		\hline
					  &JF12 & 26 & 17 & 11  & 8.33  & 6.75  & 6.4\\
					  &JF12P& 32 & 22 & 14  & 11    & 9.25  & 7.8\\
		M 82       &JF12K& 26 & 17 & 11  & 9     & 7.25  & 6.8\\
					  &PTKN & 32 & 18 & 11  & 8.33  & 7.25  & 6.2\\
					  &TF17 & 62 & 40 & 26  & 19.67 & 17.75 & 13.6\\
 		\hline
					  &JF12 & 24 & 15 & 10.5& 8     & 7     & 6.4\\
					  &JF12P& 32 & 20 & 13.5& 10.33 & 9.5   & 7.8\\
		M 87       &JF12K& 24 & 15 & 10  & 8.33  & 7     & 6.4\\
					  &PTKN & 28 & 17 & 11.5& 9     & 7.25  & 5.8\\
					  &TF17 & 60 & 38 & 28  & 19    & 16.25 & 13.8\\
 		\hline
					  &JF12 & 26 & 16 & 10  & 8.67  & 7.25  & 6.4\\
					  &JF12P& 34 & 21 & 15  & 11    & 9.5   & 8.2\\
		Fornax A   &JF12K& 26 & 17 & 11  & 8.67  & 7.5   & 7.2\\
					  &PTKN & 32 & 19 & 12  & 9     & 7.75  & 6.2\\
					  &TF17 & 62 & 38 & 29  & 19    & 18    & 15.4\\
		\hline
	\end{tabular}
\end{table}

An important point to notice about results presented in
table~\ref{table:CenA_3to1} is that efficiency of the classifier remains
high when tested on the JF12 model with the total fraction of \uhecrs\
coming from any of the four pairs of sources being in the range 15--17\%
for $N=100$ and 6.4--6.8\% for $N=500$. Nearly the same fractions are
needed if the CNNs are tested on the JF12K model. As expected, other
models of the GMF need a higher contribution of from-source events but
there is no drastic decrease of performance in comparison with the case
of a single source, cf.\ table~\ref{table:B_dependence}.

Now, let us discuss what can be done in a more general case.  It is
straightforward to train a CNN for a set of nearby sources, e.g., Cen~A,
Fornax~A and M~87, which are often considered as the most likely
candidates.  However, one needs to estimate relative fluxes of \uhecrs\
arriving from the selected sources in the given energy range.

To the best of our knowledge, there is no unequivocal solution to the
problem yet. It is argued by a number of authors that in the case of radio
AGN, the UHECR flux generated by a particular source is proportional to
its radio luminosity, see, e.g,~\cite{Caramete2015,Rachen2019}.  One can
scale fluxes from each AGN taking them proportional to their radio
luminosity at some frequency (say, at 1.4~GHz as was done in the
catalogue~\cite{vanvelzen}) divided by the square of the distance to the
source. Applying this approach, one immediately finds that Cen~A, which
is a strong and nearby radio AGN, dominates the flux from the sources
considered above, with only approximately~17\% of it attributed to
M~87 and Fornax~A.  This results in a situation very similar to the case
of Cen~A being the only source of a large-scale anisotropy with just a
slightly higher fraction of from-source events necessary to observe a
deviation from isotropy.  Another approach relates the flux of \uhecrs{}
to the gamma-ray luminosity of sources~\cite{2010ApJ...724.1366D}
(see~\cite{2018ApJ...853L..29A} for a recent application of the
approach). In this case, Cen~A also dominates the UHECR flux but in a
different proportion to the flux of other sources.

To avoid the necessity of estimating the relative flux of possible
nearby sources, we tried a more generic way.  Namely, we assumed that
any of the three sources~--- Cen~A, M~87 and Fornax~A~--- can be a
dominating one but we do not know which of them.  Then we trained
classifiers with samples created the same way as in the cases discussed
above, i.e., with a contribution from one of them and an isotropic
background, but the source was chosen from the three in a random fashion
for each new training sample.  Once again, the CNNs were trained with
the JF12 model of the GMF with the PTKN model employed for creating
validation sets in the early-stop procedure.

After training the CNNs, we tested them against the same null
hypothesis~$H_0$ of isotropy as above. Notice that training was
performed with three sources but tests were run for each of them
separately. We also tested the performance of the CNNs for two
``unknown'' sources~--- NGC~253 and M~82, which were not anyhow involved
in training.  Finally, we checked how robust are the new classifiers
against other models of the GMF.  Results demonstrating the test
statistic performance are presented in
table~\ref{table:train3test5}.\footnote{We also
	trained neural networks using all five sources in the same fashion.
	Results of the tests for Cen~A, M~87 and Fornax~A did not differ
	strongly from those shown in table~\ref{table:train3test5} but,
	as one could anticipate, the classifiers performed much better
	for NGC~253 and M~82.}

\begin{table}[!ht]

	\caption{Results of tests of the CNNs trained simultaneously for
	three sources (Cen~A, M~87 and Fornax~A) assuming only one of them is
	contributing to a large-scale anisotropy of the flux.  NGC~253 and
	M~82 were not involved in training.  Notation is the same as in
	table~\ref{table:B_dependence}.
	See figure~\ref{fig:tables} below for an illustration of the data for M~87.
	}

 	\label{table:train3test5}
 	\medskip
 	\centering
 	\begin{tabular}{|l|c|c|c|c|c|c|c|}
 		\hline
		Source & GMF& 50 & 100 & 200 & 300 & 400 & 500 \\
 		\hline
						& JF12  & 22 & 14 & 9.5  &  7.3 &  6.25&  5.6 \\
						& JF12P & 24 & 16 & 11   &  8.3 &  7   &  5.6 \\
		Cen A       & JF12K & 24 & 15 & 9.5  &  7.67&  6.5 &  5.8 \\
						& PTKN  & 22 & 14 & 9    &  7   &  5.25&  4.4 \\
						& TF17  & 42 & 27 & 18.5 &  15  &  11  &  9.4 \\
		 \hline       
						& JF12  & 28 & 19 & 12   &  11  &  9.5 &  7.8 \\
						& JF12P & 58 & 36 & 26.5 &  22  & 19.5 & 11.8 \\
      M 87        & JF12K & 30 & 20 & 13   & 11.3 &  9   &  8   \\
						& PTKN  & 32 & 20 & 12   &  9.67& 8.75 &  6   \\
						& TF17  & -- & 95 & 60.5 & 39.7 & 31.23   & 26.4 \\
		 \hline       
						& JF12  & 18 & 13 &  8   &  6.3 &  6   &  5.2 \\
						& JF12P & 32 & 19 & 13   & 10.3 &  9   &  8   \\
		Fornax A    & JF12K & 18 & 12 & 7.5  &  6   &  5   &  5   \\
						& PTKN  & 36 & 21 & 14   & 10   & 8.5  & 8.4  \\
						& TF17  & 70 & 41 & 24   & 17   & 15.5 & 15.2 \\
		\hline
		\hline
						& JF12  & 36 & 22 & 15.5 & 11.67& 10.25& 8.2  \\
						& JF12P & 76 & 48 & 31.5 & 25   & 22&15.8  \\
		NGC 253     & JF12K & 32 & 20 & 12.5 & 9.7 & 8.25  & 6.4  \\
						& PTKN  & -- & 88 & 63.5 & 48   & 41.5 & 26.8 \\
						& TF17  & -- & 75 & 61.5 & 42.3 & 34 & 29.8 \\
		\hline       
						& JF12  & 50 & 32 & 20   & 15 & 13.25& 10.4 \\
						& JF12P & 50 & 32 & 18.5 & 15.3 & 12.75& 10.2 \\
		 M 82           & JF12K & 50 & 32 & 20   & 15.3 & 12.25& 10.8 \\
						& PTKN  & 32 & 22 & 13   & 10.3 & 9  & 7.4  \\
						& TF17  & 48 & 28 & 18   & 13.7 & 12 & 10   \\
		\hline
	\end{tabular}
\end{table}

One can see the efficiency of the classifiers did not decrease
drastically for the three sources used for training in comparison with
the ``ideal'' case shown in table~\ref{table:B_dependence}, except for
M~87 and the TF17 model.  The CNNs performed pretty well in all cases
providing the test model of the GMF does not deviate strongly from the
one used for training.  The situation with the ``unknown'' sources is
more complicated and varies depending on the source and the GMF
model.
The neural networks performed pretty well for NGC~253 with the JF12
and JF12K models, especially for $N\ge200$, but not the other models of the GMF.
Surprisingly, the best results for M~82 were obtained with the PTKN model,
which outperformed all models of the JF family.

\begin{figure}[!ht]
	\figw{.48}{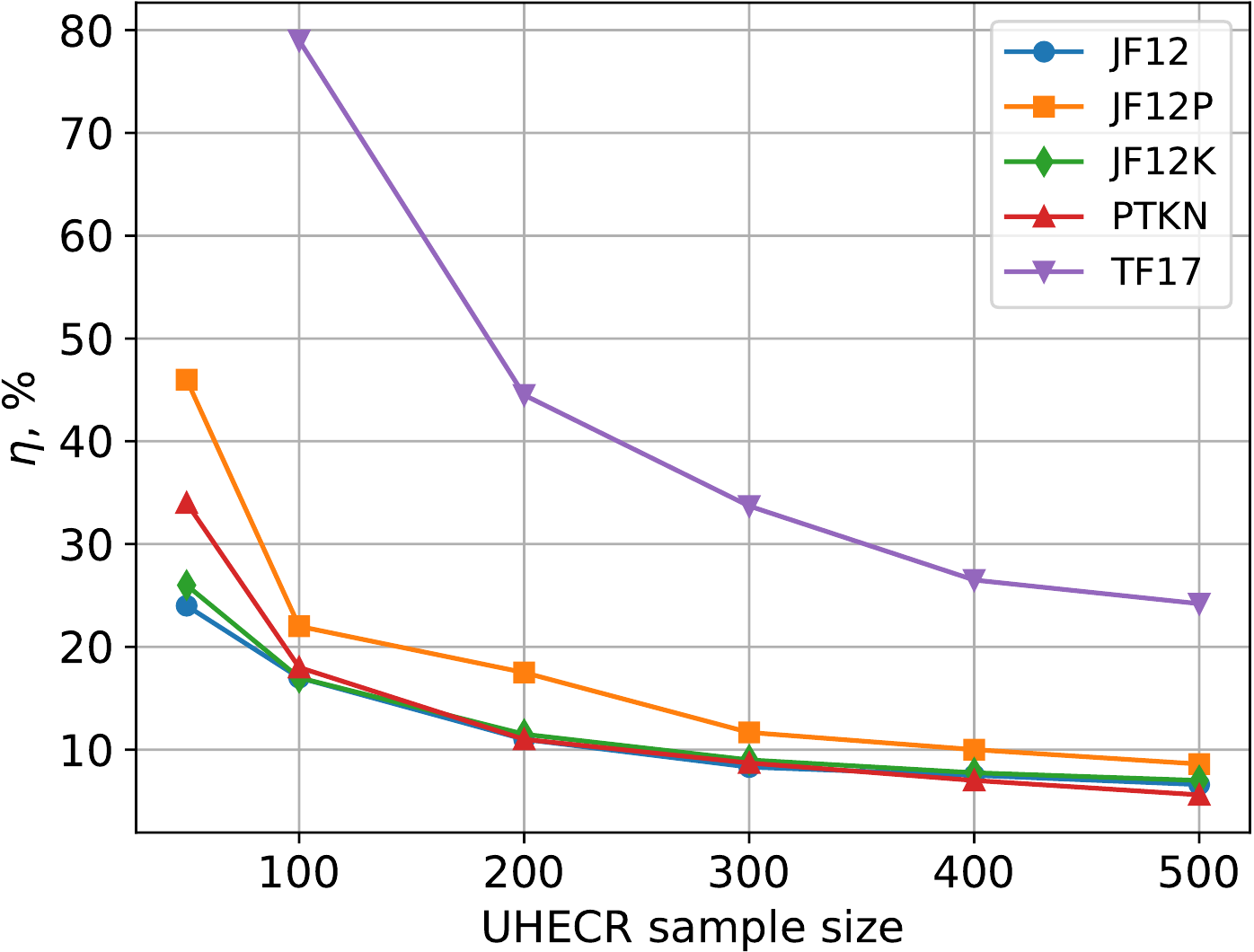}\quad
	\figw{.48}{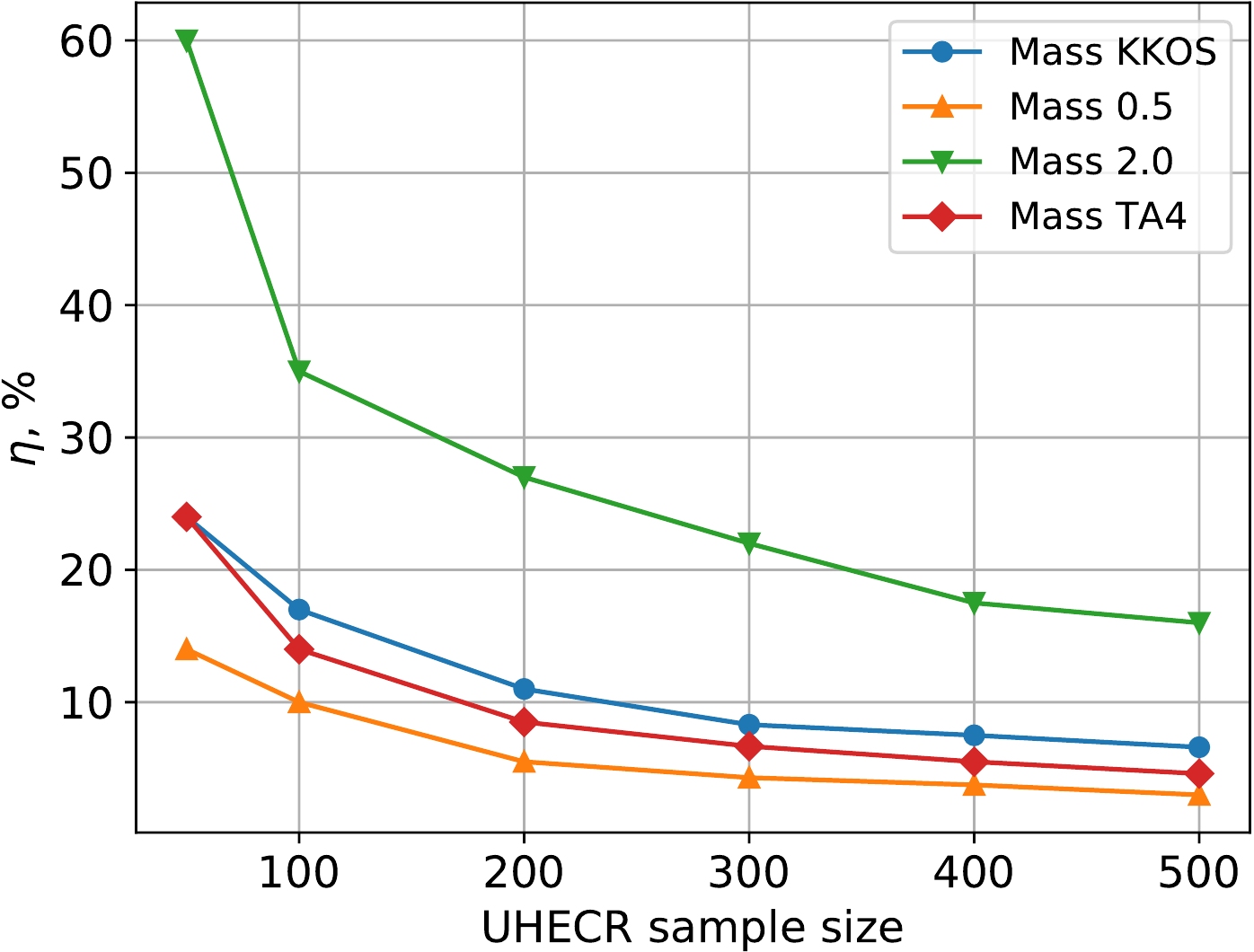}\\[2mm]
	
	\figw{.48}{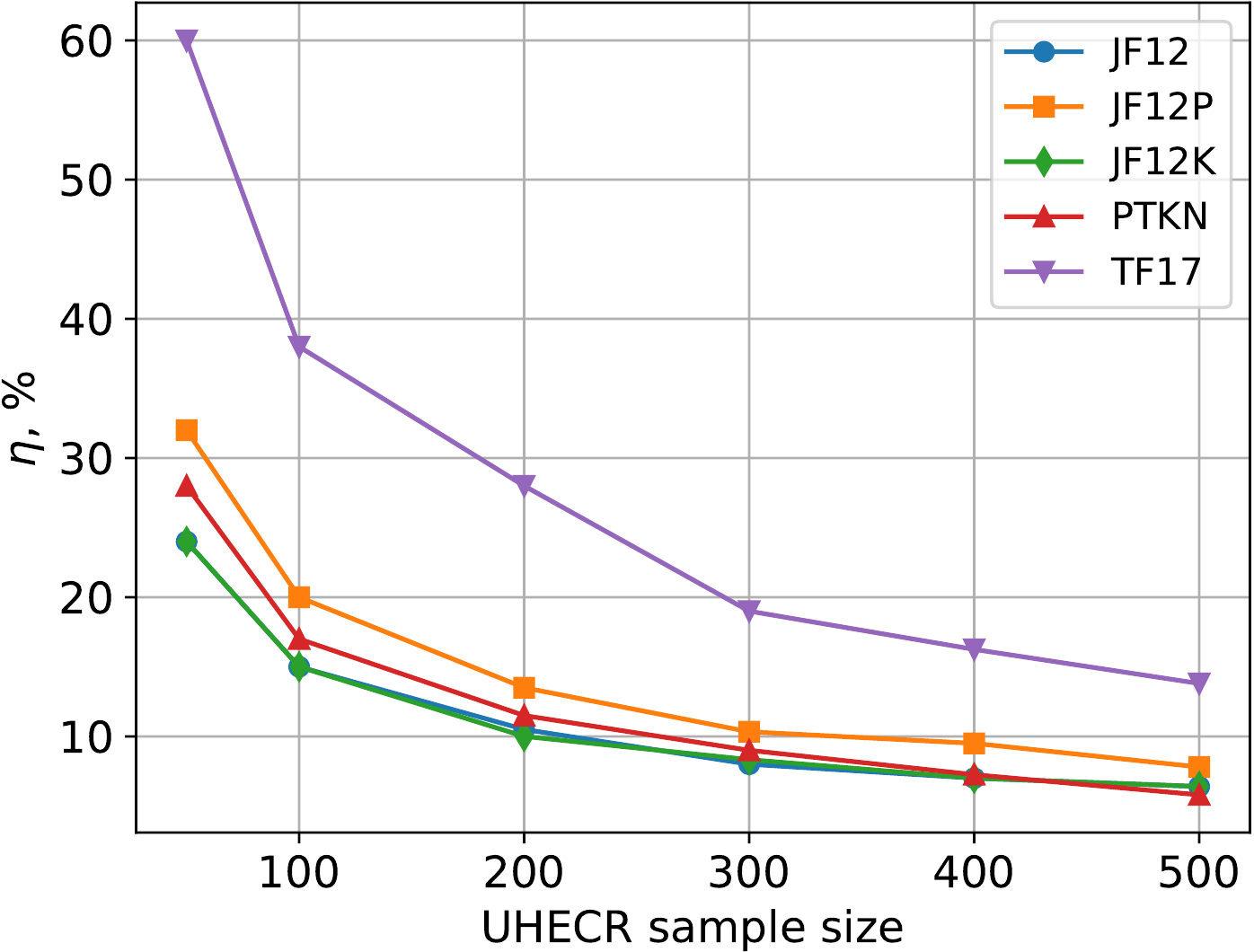}\quad
	\figw{.48}{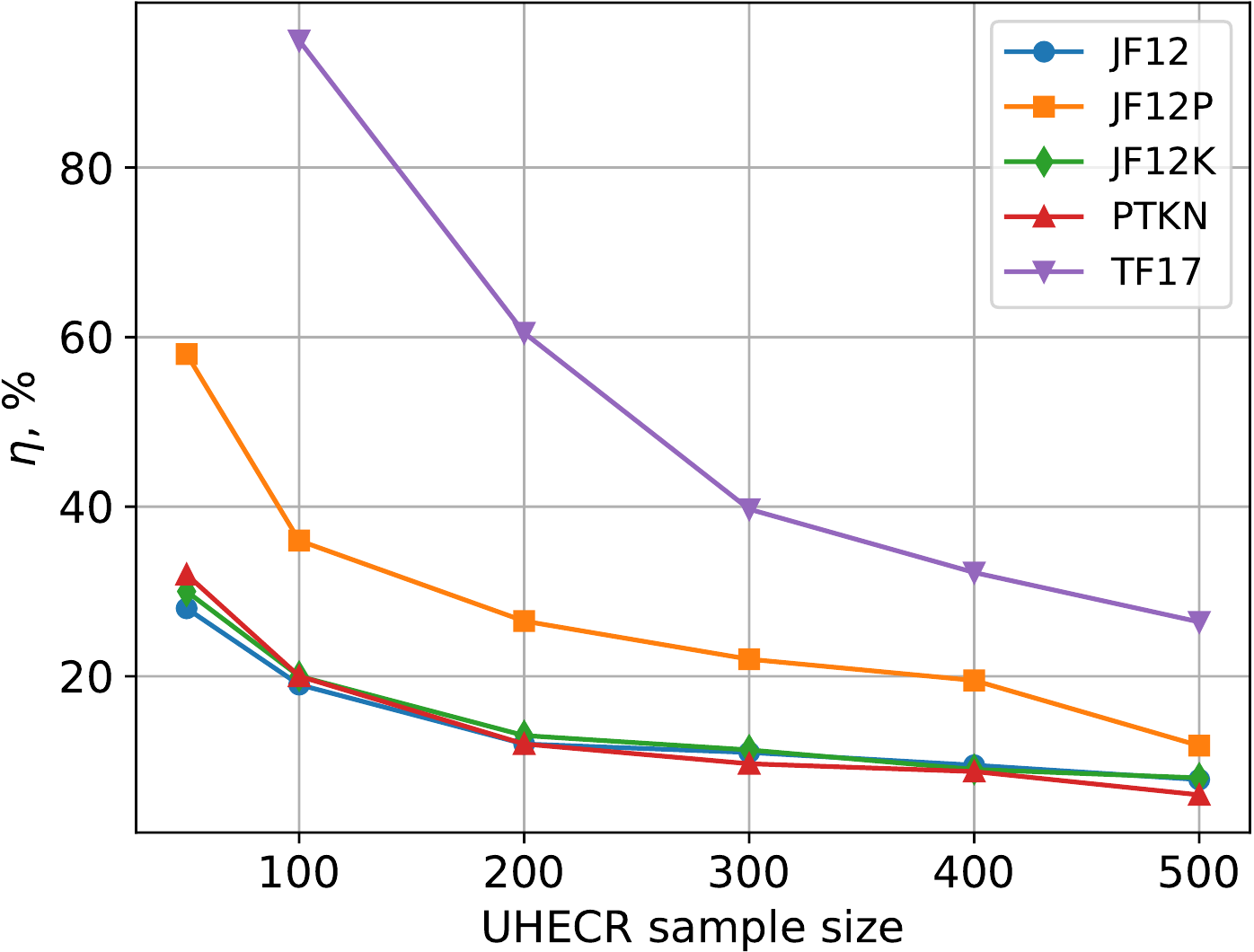}
	\caption{Illustration of the data presented in
	tables~\ref{table:B_dependence} (top left),
	\ref{table:cnn_vs_mass} (top right),
	\ref{table:CenA_3to1} (bottom left)
	and~\ref{table:train3test5} (bottom right) for M~87.}
	\label{fig:tables}
\end{figure}

Finally, figure~\ref{fig:tables} illustrates the data presented
in tables~\ref{table:B_dependence}--\ref{table:train3test5} for M~87.

%_______________________________________________________________________
\section{Conclusions}

We have demonstrated how one can strongly improve the efficiency of an
analysis of arrival directions of \uhecrs{} using machine learning
techniques. The basic idea is to train a classifier which discriminates
samples generated assuming null (isotropy) and alternative (anisotropy)
hypotheses and to use the classifier output as a test statistic.
An application of models that involve pattern recognition such as the
suggested deep convolutional neural network on a HEALPix grid gives a
really qualitative enhancement in terms of sensitivity to deviations
from an isotropic distribution of arrival directions. It was shown in
particular that the method allows decreasing the minimal number of
events necessary to reject the null hypothesis by $\sim4$ times.
This reduces technical demands and the required total exposure
of an UHECR experiment drastically.

Although the new test statistic depends on the alternative model details
such as the choice of a nearby source, the Galactic magnetic field
configuration, the spectrum and mass composition of cosmic rays, we
proposed a couple of methods to make it less model dependent. The first
idea is to use sample sets derived in a modified model as validation
data for the early-stop technique to avoid overfitting.  The second idea
is to use several sources for building the classifier training samples
to obtain a more universal test statistic.

Note that the test statistic model dependence is not necessarily a
drawback for the source identification problem. We intentionally started
our discussion from the test statistics trained with a particular source
in mind since these statistics are the most sensitive to the flux from
that source and can be used to reject the hypothesis of the particular source
giving the main contribution to the anisotropy.
As an illustration, we tried to reject the null hypothesis of Cen~A giving
at least 6\% of events, assuming that we have observed 300 events and a
large-scale anisotropy has been established as above\footnote{The fraction
of 6\% is roughly the level which allows rejecting the isotropy hypothesis,
provided that Cen~A is the strongest source (see Table~\ref{table:percentNN})}.
We needed to formulate an alternative hypothesis to do this.
We just repeated our previous calculations with null and alternative
hypothesis swapped in the case of isotropy being an alternative hypothesis.
This way, we obtained $\beta \leq 0.048$ for $\alpha = 0.01$ using the test
statistic based on the previously trained CNN classifier,
discriminating maps with an admixture of Cen~A events from isotropic maps.
A less trivial question is whether one can use the same test statistic with
another alternative hypothesis, e.g., assuming another source giving the main
contribution to the anisotropy. To answer this question, we calculated the same
Cen~A-based test statistic on maps with an admixture of 6\% events from one of
the other four sources, considered in the work.
We obtained values of~$\beta$ in the
range from~0.02 for M~82 to~0.16 for Fornax~A,
assuming $\alpha = 0.01$  as above, thus indicating the usefulness of the test
statistic derived.
However, a more efficient test statistic can be constructed for any
specific alternative hypothesis by training the CNN classifier on the
corresponding null and alternative hypotheses maps.

We emphasise the proposed method can be applied in more complicated
situations than those discussed above. For instance, we considered the
case of a uniform exposure of the celestial sphere, which is expected
for orbital detectors like \keuso{} or POEMMA but a non-uniform
exposure, which takes place for the existing ground-based experiments,
is not a problem here since it can be easily taken into account when
generating training and validation sets of data.  Besides this, one can
employ other possible sources of \uhecrs{} and their combinations to
study more cases.  Last but not least, the presented CNN is modest on
computational resources allowing one to perform the whole analysis on an
average desktop computer not necessarily equipped with a GPU.

Finally, we remark that training one CNN for each particular source and
each size~$N$ of UHECR samples was a big work. A reasonable question is
whether one can reduce the amount of calculations.  The answer is yes.
Namely, we have tested two CNNs for each of the sources (for the JF12
model), one trained for samples of the size $N=50$ and another one for
$N=500$, against the whole range of sample sizes.  A remarkable thing is
the difference between results obtained with ``dedicated'' CNNs as shown
in tables~\ref{table:percentNN} or~\ref{table:B_dependence} and those
obtained for CNNs trained for~$N=50$ or~500 was negligible.

In the future, we plan to take into account uncertainties in the energy
of detected \uhecrs{}, which influence the spectrum and thus the shape
of patterns formed by nuclei arriving from a source.\footnote{We thank
	Olivier Deligny for attracting our attention to the point.}
It is also interesting to check if a further improvement in recognizing
patterns of arrival directions and thus identifying sources of \uhecrs{}
can be obtained by changing the energy threshold of events selected for
the analysis and by incorporating information on the energy or the depth
of maximum of registered events in a neural network.  We believe the
suggested approach opens new promising possibilities for studying
anisotropy of ultra-high-energy cosmic rays and identifying their
sources. The source code and supplemental materials for this work,
including trained classifier models, can be downloaded from the project
web page~\cite{gitrepo}.

%_______________________________________________________________________
\acknowledgments

The research has made use of the NASA/IPAC Extragalactic Database (NED),
which is operated by the Jet Propulsion Laboratory, California Institute
of Technology, under contract with the National Aeronautics and Space
Administration, and of the SIMBAD database, operated at CDS, Strasbourg,
France~\cite{simbad}.  The development of the classification method and
the architecture of the corresponding deep convolutional neural network
is supported by the Russian Science Foundation grant 17-72-20291.  We
acknowledge partial financial support from the Russian Foundation for
Basic Research grant No.\ 16-29-13065.  MP acknowledges the support from
the Program of development of M.V.\ Lomonosov Moscow State University
(Leading Scientific School ``Physics of stars, relativistic objects and
galaxies'').
MZ was partially supported by the State Space Corporation ROSCOSMOS and
M.V.~Lomonosov Moscow State University through its ``Prospects for
Development'' program.
Some of the results in this paper were obtained using the
HEALPix package~\cite{healpix}.

%_______________________________________________________________________
% References
\bibliographystyle{JHEP}
\bibliography{nnaniso}

\end{document}